\documentclass[journal,lettersize]{IEEEtran}
\usepackage{graphicx}
\usepackage{algorithm}
\usepackage{algpseudocode}
\usepackage{svg}
\usepackage{float}
\usepackage{xcolor}
\usepackage{booktabs}
\usepackage{multirow}
\usepackage{makecell}
\usepackage{tabularray}
\usepackage{pifont}
\usepackage{threeparttable}
\usepackage[numbers,sort&compress]{natbib}

\newcommand{\cmark}{\ding{51}} % 打勾符号
\newcommand{\xmark}{\ding{55}} % 叉号符号

\ifCLASSINFOpdf
\else
\fi

\begin{document}
\title{A Hybrid BPMN-DMN Framework for Secure Inter-organizational Processes and Decisions Collaboration on Permissioned Blockchain}


\author{\IEEEauthorblockN{Xinzhe Shen\textsuperscript{1}, 
Jiale Luo\textsuperscript{1}, 
Hao Wang\textsuperscript{2}, 
Mingyi Liu\textsuperscript{1}, 
Schahram Dustdar\textsuperscript{3} Fellow, IEEE,
and Zhongjie Wang\textsuperscript{1}} \\
\IEEEauthorblockA{
Email: \{xzshen, jiale.luo, 2022211249\}@stu.hit.edu.cn, 
\{liumy, rainy\}@hit.edu.cn},
{dustdar@dsg.tuwien.ac.at}\\
\IEEEauthorblockA{\textsuperscript{1}\textit{Faculty of Computing, Harbin Institute of Technology, Harbin, China}} \\
\IEEEauthorblockA{\textsuperscript{2}\textit{Faculty of Computing, Harbin Institute of Technology, Weihai, China}}\\
\IEEEauthorblockA{\textsuperscript{3}\textit{Distributed Systems Group, TU Wien}}%
\thanks{Manuscript received December 1, 2015; revised August 26, 2015. Corresponding author: Mingyi Liu (email: liumy@hit.edu.cn).}}

\markboth{Journal of \LaTeX\ Class Files,~Vol.~14, No.~8, August~2024}%
{Shell \MakeLowercase{\textit{et al.}}: Bare Demo of IEEEtran.cls for IEEE Transactions on Magnetics Journals}

\IEEEtitleabstractindextext{%
\begin{abstract}
In the rapidly evolving digital business landscape, organizations increasingly need to collaborate across boundaries to achieve complex business objectives, requiring both efficient process coordination and flexible decision-making capabilities. Traditional collaboration approaches face significant challenges in transparency, trust, and decision flexibility, while existing blockchain-based solutions primarily focus on process execution without addressing the integrated decision-making needs of collaborative enterprises. This paper proposes BlockCollab, a novel model-driven framework that seamlessly integrates Business Process Model and Notation (BPMN) with Decision Model and Notation (DMN) to standardize and implement collaborative business processes and decisions on permissioned blockchain platforms. Our approach proposes a multi-party collaboration lifecycle supported by BlockCollab, and automatically translates integrated BPMN-DMN models into smart contracts(SCs) compatible with Hyperledger Fabric, enabling privacy-aware multi-organizational process execution through blockchain-based Attribute-Based Access Control (ABAC). The framework introduces three key innovations: (1) a standardized method for modeling collaborative processes and decisions using integrated BPMN-DMN model, (2) an automated SC generator that preserves both process logic and decision rules while maintaining privacy constraints, and (3) a hybrid on-chain/off-chain execution environment that optimizes collaborative workflows through secure data transfer and external system integration. Experimental evaluation across 11 real-world collaboration scenarios demonstrates that our approach achieves 100\% accuracy in process execution. Furthermore, an analysis of various execution processes highlights the strong practical applicability and reliability of our approach. The proposed framework includes an open-source\footnote{https://github.com/XinzheShen182/ChainCollab} third-party collaboration platform based on blockchain.
% Security analysis confirms robust privacy protection through ABAC and the hybrid architecture.

\end{abstract}

}

\maketitle

\IEEEdisplaynontitleabstractindextext
\IEEEpeerreviewmaketitle

\section{Introduction}\label{section:intro}
\subsection{Background}
 
\IEEEPARstart{I}{n} the rapidly evolving landscape of contemporary business, collaboration has emerged as a fundamental pillar of organizational success. The intricacies of modern economic activities render it impossible for any single organization to autonomously address all business requirements. Inter-organizational collaboration facilitates the efficient integration of resources, bolsters competitive advantages, and catalyzes the development of innovative business models.

 process in which each organization participates and completes its tasks to form the entire workflow.\textit{Business processes} and \textit{decisions} are two pivotal elements\cite{OMGDMN}. \textit{Business processes} delineate the specific steps of each organization must execute within the collaborative framework to achieve shared business objectives. A business process consists of multiple activities, 
a subset of which represents decisions~\cite{haarmann2019executing_dmn}. The prpgression of the process is contingent upon the outputs of these decisions. \textit{Business decisions} in organizational collaborations transcend the boundaries of individual entities. While modifying business processes incurs substantial costs and thus occurs infrequently, business decisions exhibit a higher degree of volatility, frequently adapting in response to myriad internal and external factors.

\begin{figure}[!h]
    \centering
	\includegraphics[width=\linewidth]{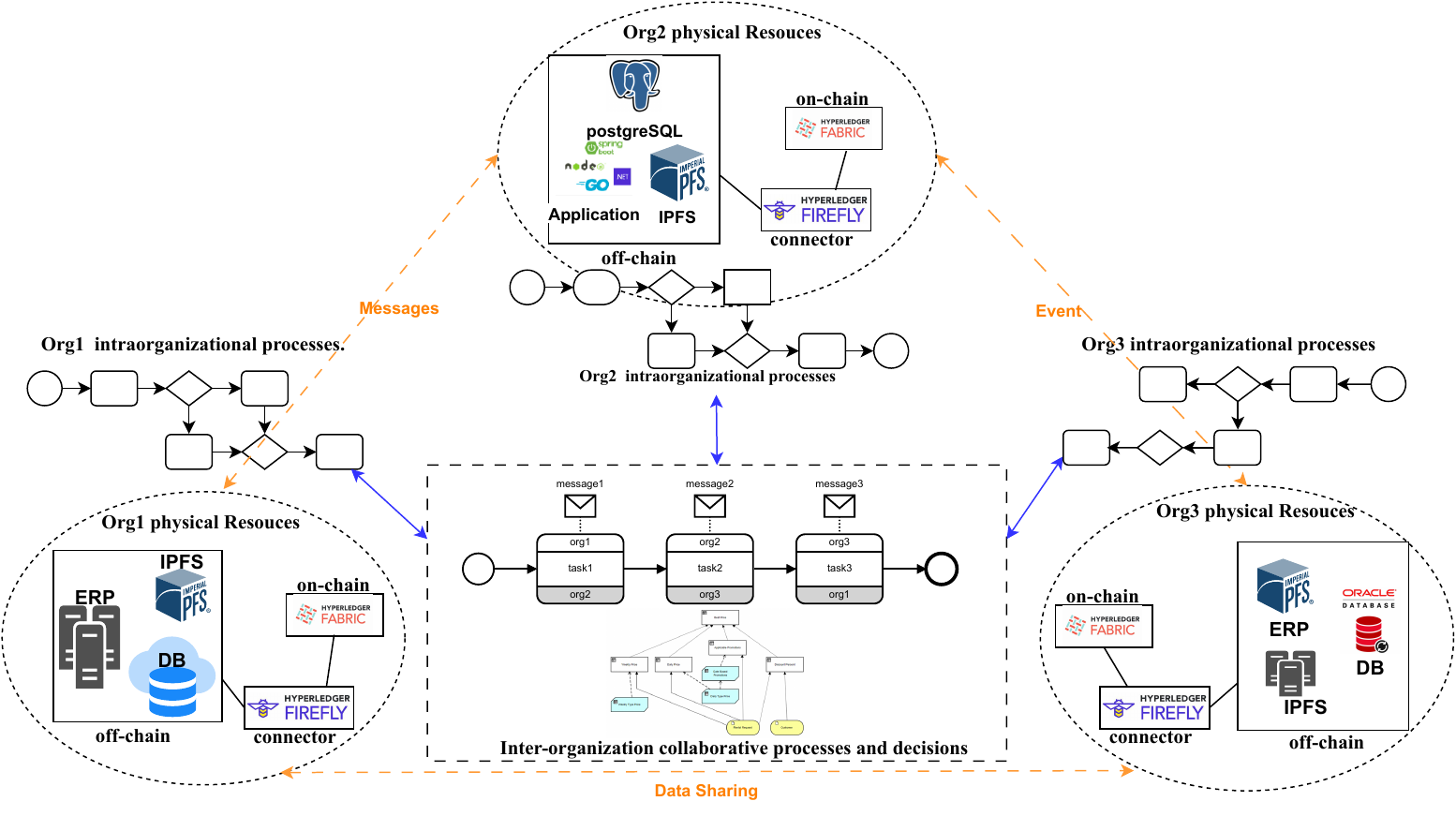}
	\caption{Inter-organizational collaboration with Blockchain}
	\label{fig:background}
\end{figure}

Organizational collaboration traditionally operates in two main modes: centralized and distributed, each significantly influencing process execution and decision-making across collaborating entities.  A central authority orchestrates business processes and heavily influences key decisions, establishing collaborative processes and significantly impacting the entire network. Conversely, the distributed mode allows for more autonomous process management and decentralized decision-making by individual organizations, aligning with collaborative goals. However, both collaboration modes exhibit inherent limitations that affect process execution and decision-making. In the centralized mode, the central authority may be susceptible to power abuse, potentially exhibiting favoritism towards certain entities, thereby undermining the equilibrium and equity of the collaborative ecosystem. In the distributed mode, while mitigating centralized control issues, presents challenges in collaboration efficacy. Participants' inability to observe the actions of their counterparts can engender communication barriers and decision-making latencies, impacting the overall business process flow.

The blockchain technology has introduced a novel mode of inter-organizational collaboration, fundamentally altering the landscape of business interactions. In this mode, participants operate without reliance on a centralized authority, benefiting from unprecedented transparency throughout the collaborative process. It also provides an immutable record of decision-making activities among organizations~\cite{lauster2020literature_survey_BCandBPMN}. Blockchain implements a decentralized trust mechanism through its distributed ledger and consensus algorithms. This architecture enables participants to independently verify and irreversibly record transactions, ensuring data transparency and integrity~\cite{weber_choreography}. Furthermore, blockchain's chaincode can facilitate the encoding of collaborative processes and decision logic into self-executing code, triggered under pre-defined conditions to significantly enhancing the efficiency of inter-organizational workflows.

As shown in Fig.~\ref{fig:background}, this is a blockchain-based inter-organizational collaboration approach, supporting distributed business processes and decisions. A collaborative ecosystem comprising three distinct organizations, united by a shared business goal. Within this ecosystem, each organization autonomously executes its private business processes~\cite{inter_privateProcess} and manages its heterogeneous physical resources. The inter-organizational interaction is facilitated through blockchain-supported public collaborative processes and decision-making mechanisms. This architectural approach effectively dismantles traditional organizational data silos, resulting in a marked enhancement of collaboration efficiency and a significant increase in operational transparency.

\subsection{Motivation}
While blockchain technology offers a robust infrastructure for inter-organizational collaboration, effectively modeling and managing collaborative processes and decisions within this new mode presents its own set of challenges. To leverage the full potential of blockchain in organizational collaboration, it is crucial to have a standardized method for describing and designing collaborative business processes. This is where Business Process Modeling and Notation (BPMN)~\cite{OMG2013BPMN} comes into play, serving as a bridge between the conceptual understanding of collaborative processes and their implementation on blockchain platforms. \textbf{The integration of blockchain technology with BPMN-modeled processes has introduced new paradigms of security and transparency in executing interacting business processes }\cite{haarmann2018dmn_execution}. 

Complementing BPMN, the Decision Model and Notation (DMN)~\cite{OMGDMN} offers a standardized methodology for modeling decisions, their requirements, and dependencies, distinct from the processes flows~\cite{haarmann2018dmn_execution}. This separation of concerns has gained traction in traditional process execution environments, with established BPMN engines such as Camunda \cite{fernandez2013camunda} and Activiti \cite{rademakers2012activiti} integrating DMN as their decision modeling framework to automate decision processes. Combining BPMN with a decision engine facilitates a clear delineation between business processes and decisions. This segregation confers significant advantages: it allows for agile modifications to decision logic in response to evolving business conditions without necessitating alterations to the underlying process structure. Furthermore, it simplifies BPMN process modeling by eliminating the need for redundant gateway constructs. However, \textbf{in the context of blockchain-based process execution, there is a notable absence of methodologies for implementing DMN-modeled decisions in conjunction with BPMN processes.}

To implement the aforementioned process and decision in a blockchain environment, developing a Web3.0 Decentralized Application (DApp) is essential. DApps have become the primary solution for achieving multi-authority business processes in this new web era~\cite{MDAPW3}. However, \textbf{the complex architecture of decentralized DApps, including on-chain, off-chain, and hybrid on-chain/off-chain components (connectors), makes their development challenging}~\cite{MDAPW3}. Therefore, a model-driven approach (MDA) is crucial for constructing products compatible with blockchain platforms~\cite{MDAPW3}. The primary goal of MDA is to achieve interoperability across tools and establish the long-term standardization of models in popular application domains~\cite{omg_mda_guide}. This approach can guide us in standardizing organization collaboration models, simplifying SC development and the setup of on-chain and off-chain environments.
Additionally, the integration also presents novel challenges, particularly in translating BPMN models into blockchain-compatible implementations. Recent studies have explored the generation of SC code from BPMN choreography or collaboration diagrams, aiming to ensure trustworthy execution of collaborative processes 
\cite{weber_choreography,corradini2021model_multiBlockchain,Traceability,Dynamic-Integration,chor-chain,Efficient_renda}. These studies primarily employ a model-driven approach to generate SC code that supports blockchain execution, thereby automating processes, enhancing developer efficiency, and reducing the risk of errors introduced by software developers. Despite these advancements, a significant limitation in the current  body of literature is the predominant focus on permissionless blockchain platforms (e.g., Ethereum \cite{Ethereum}). Comparatively little attention has been given to permissioned blockchains (e.g., Hyperledger Fabric~\cite{fabric}), which offer superior privacy features and are generally more suitable for organizational collaborations. Consequently, the additional complexity introduced by the underlying infrastructure of permissioned blockchains in SC development, especially when derived from BPMN models, has been largely overlooked \cite{bodorik2023_TABS}. This gap in research presents an opportunity to explore how BPMN can be more effectively utilized in the context of permissioned blockchain-based collaborative processes.

\begin{figure*}
    \centering
    \includegraphics[width=0.9\linewidth]{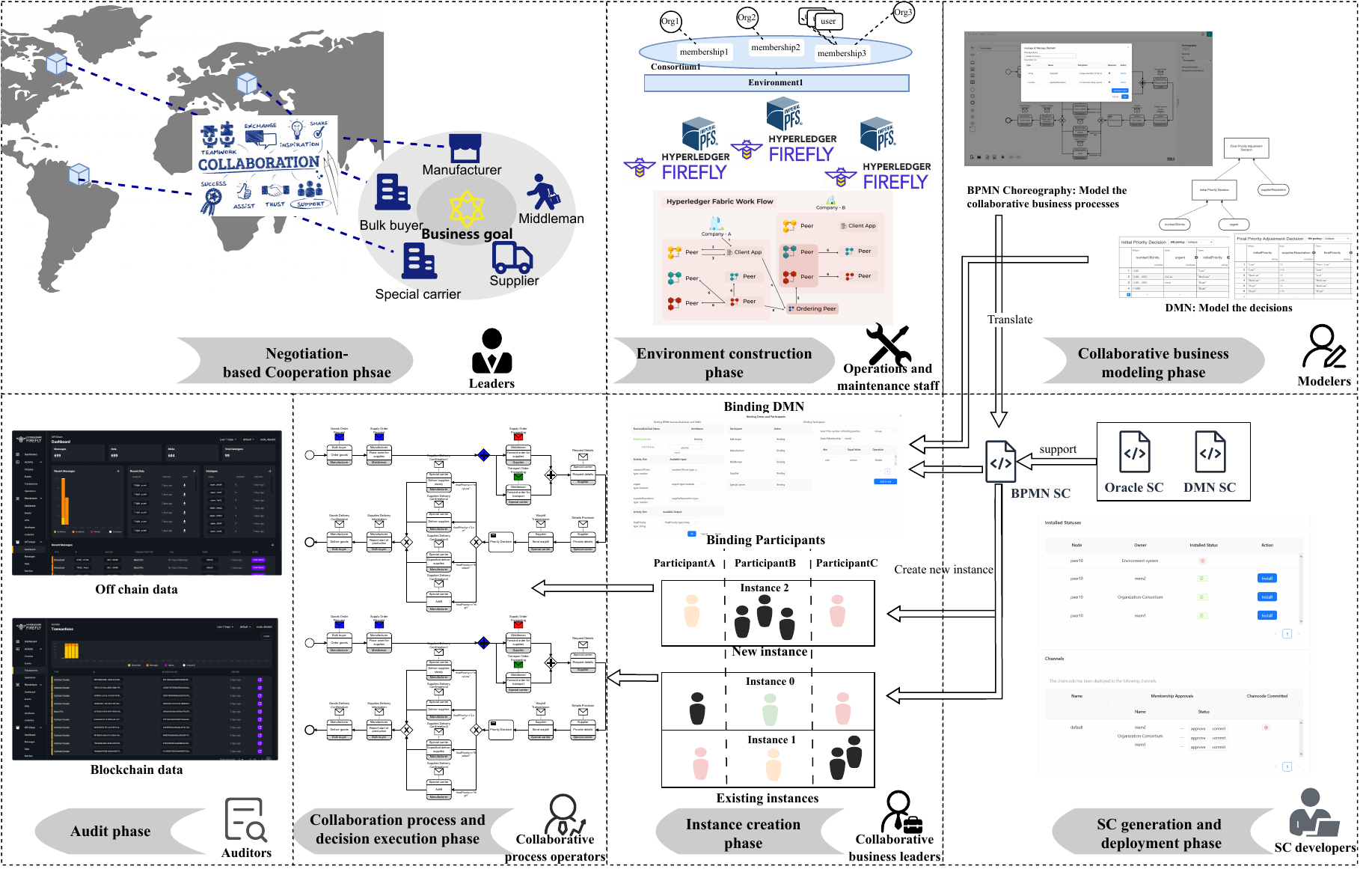}
    \caption{The lifecycle for multi-party collaboration supported by BlockCollab}
    \label{fig:life-cycle}
\end{figure*}

\subsection{Contributions}
This work presents the BlockCollab model-driven framework and its corresponding lifecycle for multi-party collaboration(as depicted in Fig.~\ref{fig:life-cycle}), which integrates collaborative processes, decision-making, and blockchain technology. It utilizes a model-driven approach to model collaborative processes and decisions, generate SC code, and design a heterogeneous on-chain and off-chain environment architecture. The key contributions of this paper are summarized as follows:

\begin{enumerate}
    \item \textbf{A standardized business collaboration modeling method} that integrates DMN with the BPMN choreography model for modeling business processes and decisions in multi-organizational collaborations. This addresses the need for a standardized method to describe and design collaborative business processes and decisions in a blockchain context.
    \item A \textbf{SC Translator} translates integrated BPMN-DMN business models into Hyperledger Fabric SC code, enabling the execution of multi-instance collaborative business processes and decision execution. At the same time, managing collaborative identities based on blockchain Attribute-Based Access Control (ABAC).

    \item An extended method based on~\cite{shenIcws} is proposed for constructing \textbf{an innovative hybrid on-chain and off-chain execution environment}. The hybrid environment provides: 1) a collaboration model to ensure participants clearly understand their roles and map physical resources to optimize the construction of the environment; 2) a blockchain-based on-chain and off-chain environment using Hyperledger Fabric and InterPlanetary File System (IPFS); 3) a connector that links on-chain and off-chain systems, supporting smooth integration with external systems; 4) Integration with an Oracle, bridging the gap between reality and blockchain.
    % utilize \textit{Private data bus} and InterPlanetary File System (IPFS) for secure off-chain data transfer and on-chain verification; 3) smooth integration with external systems through \textit{SC APIs} and \textit{Event bus}.

    \item  A fully \textbf{open-source blockchain-based third-party collaboration platform} that incorporates the proposed methods and environments. 
\end{enumerate}

\section{Method}

\subsection{Lifecycle of the Blockchain-Based Collaboration}

This section presents a comprehensive lifecycle for multi-party collaboration, integrating collaborative processes, decision-making, and blockchain technology. The lifecycle consists of seven stages (as illustrated in Fig.~\ref{fig:life-cycle}), each addressing specific aspects of collaboration implementation and execution. 

The \textbf{\textit{Negotiation-Based Collaboration phase}} initiates the collaborative journey, typically driven by one or several organizations within the supply chain. During this phase, organizational leaders and strategic decision-makers engage in comprehensive negotiations to establish the foundation for collaboration. They focus on defining clear roles and responsibilities, establishing governance frameworks, and determining benefit distribution mechanisms across participating organizations. This critical phase culminates in formal agreements that outline collaboration objectives, success metrics, and operational parameters, ensuring all parties have a shared understanding of their commitments and expected outcomes.

The \textbf{\textit{Environment Construction phase}} (Sect.~\ref{subsection-environment}) focuses on establishing the technical infrastructure necessary for blockchain-based collaboration. Operations and maintenance staff from each organization  collaboratively create a distributed environment, starting with the setup of collaborative identities on the third-party platform, allowing organizations to initiate a consortium and invite other participants.  Subsequently, technical teams configure the Hyperledger Fabric environment, concluding with the installation of essential SCs and establishing off-chain infrastructure, comprising IPFS clusters and Hyperledger Firefly connectors.

During the \textbf{\textit{Collaborative Business Modeling phase}} (Sect.~\ref{subsection:model-method}), business process modelers and domain experts from participating organizations work together to transform the negotiated agreements into formal process and decision models. They utilize the platform's BPMN choreography tools to design collaborative processes and DMN tools to model decision rules. These models undergo thorough validation with stakeholders to ensure they accurately reflect the agreed-upon collaboration parameters. Once all organizations accept the models, they are uploaded to the platform, serving as the foundation for subsequent implementation phases.

The \textbf{\textit{SC Generation and Deployment phase}} (Sect.~\ref{sect:SC-ganerate}) transforms the business models into executable SCs. SC developers utilize the \textit{SC Translator} component to convert the integrated BPMN choreography and DMN models into suitable Fabric SC(chaincode). Following Fabric's guidelines, organizations deploy these SCs according to predefined endorsement rules, typically following a majority consensus model. This deployment process includes packaging, installation, approval, and commitment operations, with each organization's approval of the installed SCs signifying their acceptance of the contract's content and readiness for automated execution.

The \textbf{\textit{Instance Creation phase}} enables the practical application of the choreography model through multiple instances, enhancing model reusability. Business leaders from each organization can create various instances of the BPMN choreography after its transformation into SCs. Each instance requires careful configuration, including participant binding and DMN association. The participant binding can specify either individual participants or groups meeting certain ABAC conditions, while DMN binding must conform to the BRT-defined inputs and outputs specified in the BPMN. This flexibility allows for process consistency while accommodating variations in participants and decision rules, supporting the dynamic nature of business operations.

During the \textbf{\textit{Collaboration Process and Decision Execution phase}}, process operators and business users actively engage in executing the configured instances. The platform's UI interface provides real-time visibility into process progress, allowing participants to input message content directly through the platform. Organizations can also integrate SC APIs with their internal ERP systems to streamline process advancement. This phase represents the operational heart of the framework, where the designed collaboration actually takes place and delivers business value.

During the \textbf{\textit{Audit phase}}, auditors can utilize the query functionality provided by Firefly to access data stored both on the blockchain and off-chain in IPFS, enabling the review of data related to each process instance and decision.

\subsection{Standardized Business Collaboration Modeling Method}\label{subsection:model-method}
This section presents a standardized method for modeling collaborative processes and decisions using an integrated BPMN-DMN approach(Sect.~\ref{subsubsect:bpmn-elements},~\ref{subsubsect:dmn}). A supply chain scenario is provided as an example to demonstrate the example of this model(Sect.~\ref{subsubsect:case}).
\subsubsection{\textbf{BPMN choreography modeling elements}}\label{subsubsect:bpmn-elements}

BPMN diagrams are widely recognized as the standard for modeling processes across organizations~\cite{Efficient_renda}, aimed at being comprehensible to a diverse group of business stakeholders including business analysts, technical developers, and process managers~\cite{bodorik2023_TABS}. 

\begin{figure}
    \centering
    \includegraphics[width=\linewidth]{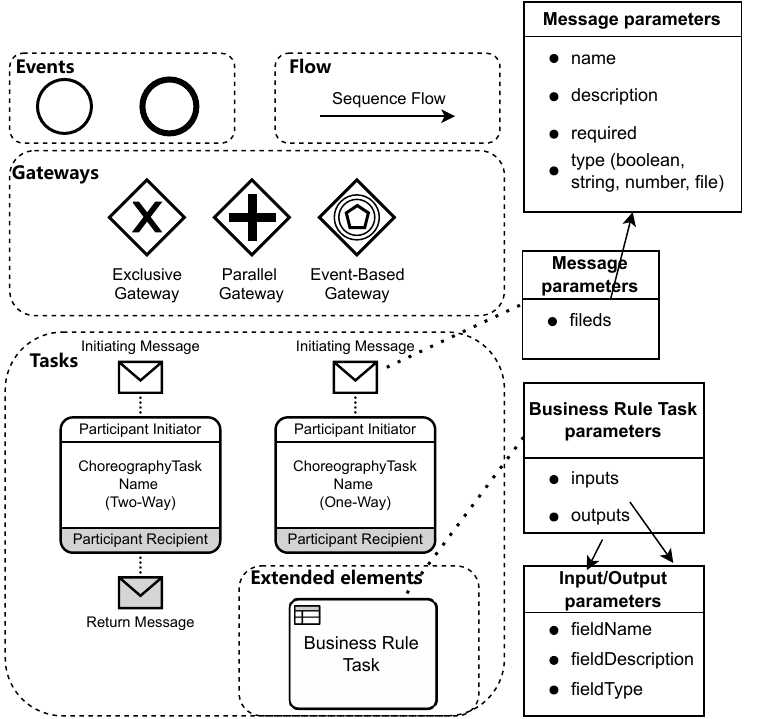}
	\caption{BPMN choreography modeling elements}
	\label{fig:bpmn-elements}
\end{figure}

An introduction to the four types of BPMN diagrams is provided in Appendix A-A. The Choreography diagram is chosen for its effective representation of autonomous and egalitarian interaction patterns among organizations. The focus is on the sequence of message exchanges and interactions among multiple participants, rather than processes controlled by a single entity. Therefore, this paper focuses on \textbf{the most common elements in \textit{Choreography diagram}} to describe the collaborative processes among multiple organizations based on blockchain technology, these elements are shown in the dotted line part of the Fig.~\ref{fig:bpmn-elements}. All interactions are explicitly recorded, allowing each participant to view the complete process state and history, thereby establishing trust across organizations. Additionally, this paper introduces the \textbf{\textit{Business Rule Task (BRT)}} from the collaboration diagram to represent decisions as activities in the process. 

\begin{figure*}[h]
    \centering
	\includegraphics[width=0.9\linewidth]{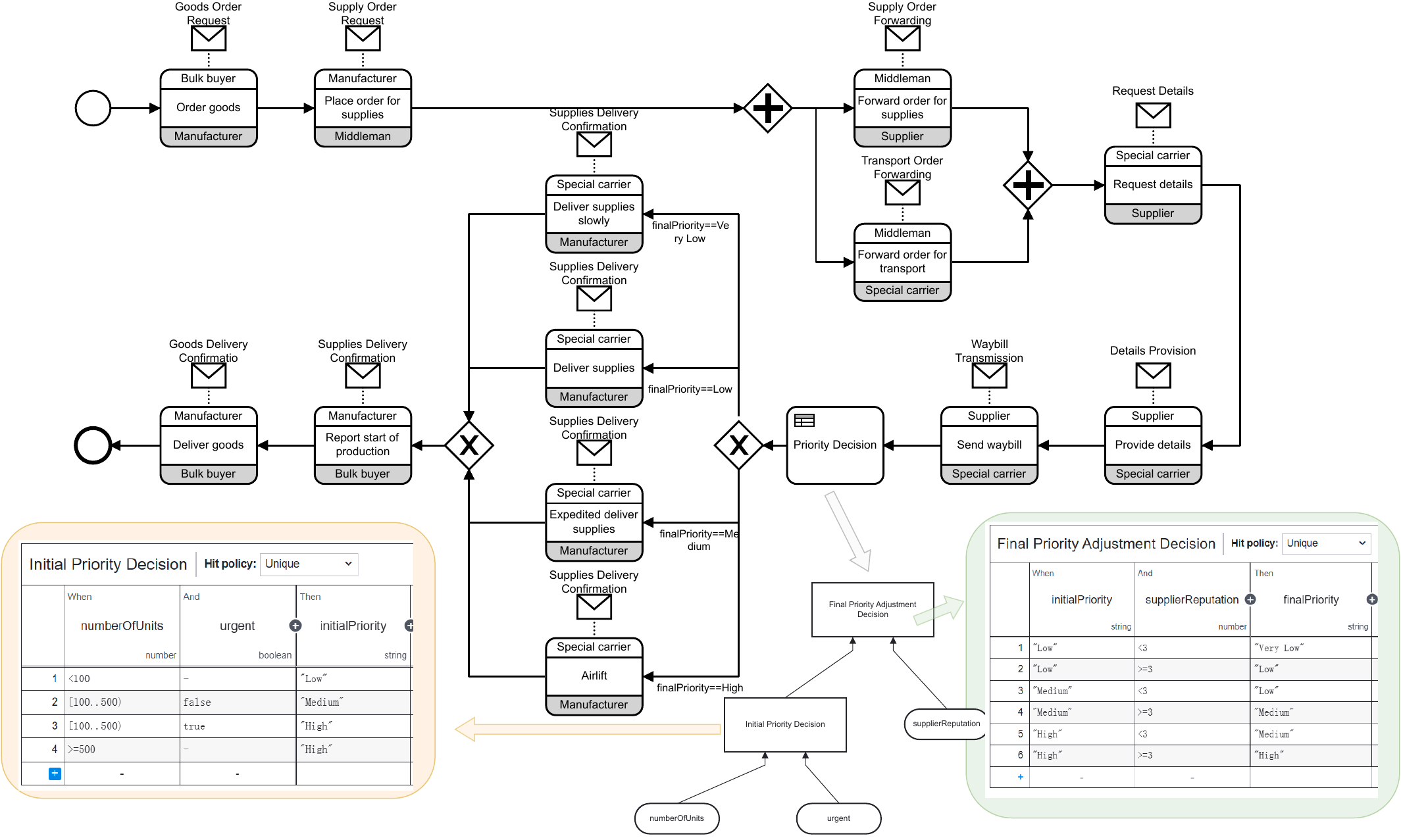}
	\caption{Example of Modeling a Supply Chain Scenario Using Integrated BPMN-DMN Models}
	\label{fig:bpmn-running-example}
\end{figure*}

The definitions of \textit{\textbf{Events, Flow, Gateways, and Tasks}} have been elaborated in the previously paper~\cite{shenIcws} and are not reiterated here for brevity. The definition of \textit{\textbf{message}} used in this paper follows that in our earlier work, as shown in the upper right corner of Fig.~\ref{fig:bpmn-elements}. Each message contains multiple fields, where each field's parameter defining its name, description, type and whether it is required. Supported types include basic JSON types (boolean, string, number) and file. The type of each field in a message is validated during runtime to ensure data conformance to the specified type.

A BRT provides a mechanism for the process to provide input to a business rules engine and to get the output of calculations that the business rules engine might provide~\cite{OMG2013BPMN}. 
A BRT is linked to a DMN Decision Requirements Diagram (DRD), which illustrates how key elements of decision making, such as domains of business knowledge, sources of business knowledge, input data, and decisions, are interconnected within a dependency network. The corresponding DRD can be binded with the BRT at the instance creation phase of a BPMN choreography diagram, rather than at the modeling stage, to support the upgrade and replacement of runtime decisions. However, when modeling BPMN choreography, it is necessary to specify the input and output data formats of a BRT to ensure that the replacement and upgrade of decisions comply with the flow of the BPMN diagram; otherwise, it will affect the subsequent process activities based on the output of the decision.

\subsubsection{\textbf{DMN}}\label{subsection:dmn}\label{subsubsect:dmn}
While DMN is essentially independent of BPMN and can function separately, it is also compatible for combined use. In our approach, we integrate BPMN with DMN to enable automated decision-making. One method for achieving automation is through “decision services (DSs),” which are deployed from a Business Rules Management System and invoked by a Business Process Management System (BPMS). These DSs encapsulate DMN-supported decision logic and provide interfaces that correspond to subsets of inputs and decisions within DMN. When invoked with a set of input data, the DS evaluates the specified decisions and returns the outputs. To implement this, we have designed a Java chaincode (SC) as DS that runs Camunda's DMN execution engine \footnote{https://github.com/camunda/camunda-engine-dmn}.

The key DMN elements adopted in this study and meaning, as shown in the Appendix Sect. A-B, focus on specifically \textit{Decision} and \textit{Input Data}, to enable automated decision-making within multi-party organizational collaboration processes. Only the essential DMN elements are implemented, as the primary requirement is to ensure reliable execution of decision rules based on data generated during collaboration, with outcomes designed to be accepted by all involved parties.

The entire DRD input is provided by a BRT, which also receives and records the decision results in the process data. As shown in the lower right corner of Fig.~\ref{fig:bpmn-running-example}, a BRT must specify multiple inputs and a single output. The input data serve as inputs to the DMN execution engine and are derived from the message data of the preceding choreography task by selecting a specific field from the message. Such data, classified as public decision data (visible to all parties in the collaborative choreography process), is automatically recorded in the SC and is no longer private (restricted visible to the sender and receiver). The output also requires specifications for its name, description, and type, as the following gateway branches will determine the execution path based on the output values.
\begin{figure*}[h]
    \centering
	\includegraphics[width=0.9\linewidth]{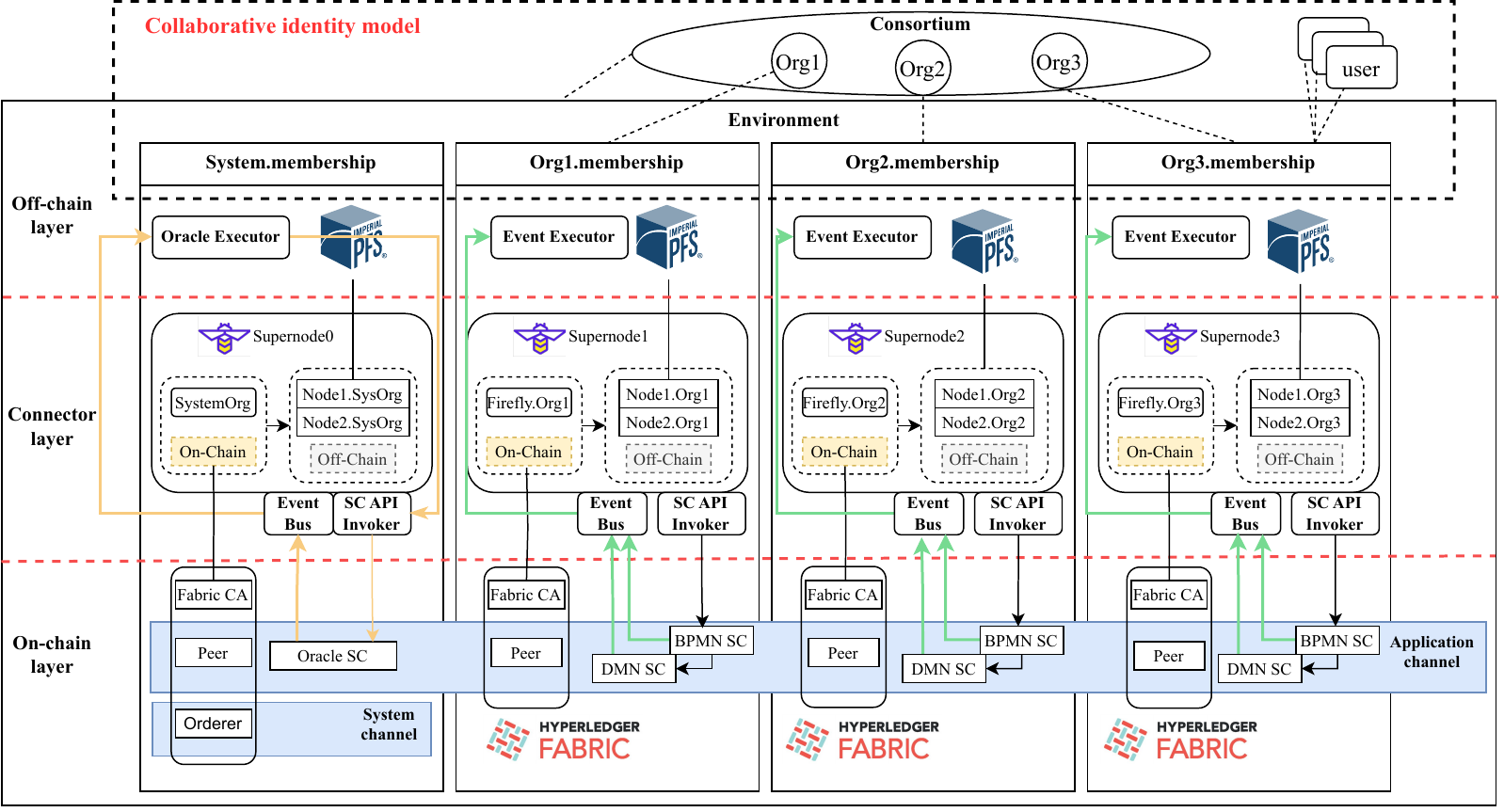}
	\caption{Hybrid On-Chain and Off-Chain Environment}
	\label{fig:physical-resources}
\end{figure*}

\subsubsection{\textbf{Case}}\label{subsubsect:case}

As shown in Fig.~\ref{fig:bpmn-running-example}, this scenario describes a supply chain involving multiple organization, further expanded from requirements extracted from the paper~\cite{supplyExample}. The BPMN choreography diagram displays a complex supply chain management process involving multiple organizations, including \textit{Bulk Buyer, Manufacturer, Middleman, Supplier, and Special Carrier}, who collaboratively complete a business process for product exchange. The process initiates when the \textit{Bulk Buyer} places an order for goods, triggering a sequence of coordinated activities across the supply chain. Upon receiving the order, the \textit{Manufacturer} requests supplies from the \textit{Middleman}, who then orchestrates the forwarding of supply and transport orders to the \textit{Supplier} and \textit{Special Carrier}, respectively.

In the BPMN choreography, the "Priority Decision" BRT follows the Special Carrier's request for details from the Supplier. This task involves assessing the urgency and volume of the order to determine the priority of transport. Additionally, the priority is adjusted based on the Supplier's reputation. The details of this DMN are shown in Fig.~\ref{fig:bpmn-running-example}. This DMN includes two Decisions, where the "Initial Priority Decision" acts as a sub-decision to the "Final Priority Adjustment Decision". The final output of the DMN, the transport priority, is then fed back into the choreography diagram to guide the process advancement.

\subsection{Hybrid On-Chain and Off-Chain Environment}\label{subsection-environment}
This article supports a hybrid on-chain/off-chain execution environment to support the operation of generated SCs, which includes the following components: (1) A collaborative identity model based on organization, consortium, membership, user and environment. (2) A physical resource construction for on-chain and off-chain environment. (3) A connector built on Hyperledger Firefly~\cite{firefly}. (4) Oracle that connects reality with blockchain

\subsubsection{\textbf{Collaborative identity model}}
To ensure that each participant in the multi-party collaboration process clearly understands their role and responsibilities, and to enable the mapping of physical resources when constructing the collaborative environment based on blockchain, we propose the collaboration identity model shown in the upper corner of Fig.~\ref{fig:physical-resources}. Our previous work~\cite{shenIcws} introduced the concepts of \textit{Organization, Consortium, Membership, and Environment} entities. In this paper, we add the \textit{User} entity, a more granular concept within each \textit{Membership}, allowing for precise identification down to individual participants. 
 
We illustrate the above concepts through a specific example shown in Fig.~\ref{fig:bpmn-running-example}. An \textit{Organization} refers to an independent entity or institution, such as Toyota. Toyota as a manufacturer connects various upstream and downstream organizations to form a \textit{Consortium} aimed at achieving shared business objectives through collaboration. A \textit{Membership} is the identity of an \textit{Organization} within a \textit{Consortium}, representing its status and role there. An \textit{Organization} may hold different \textit{Memberships} across various \textit{Consortium}. For example, Toyota acts as a Manufacturer in consortium1 but holds a Buyer membership in consortium2. \textit{Environment} is a hybrid on-chain/off-chain setup built to enable trusted collaboration through blockchain.
This section primarily introduces the design supports a DApp environment for multi-party collaborative operations, including on-chain and off-chain physical resources. 

\textbf{\textit{For the on-chain layer}}, we deploy Hyperledger Fabric, an enterprise-grade blockchain platform that enables secure and customizable collaboration~\cite{fabric}. The Fabric network architecture consists of two main types of nodes: \textit{Peer nodes} and \textit{Orderer nodes}. \textit{Peer nodes} are responsible for storing blockchain data, executing SC(chaincode) logic, and validating transaction proposals, while \textit{Orderer nodes} sequence validated transactions into blocks and broadcast them to all \textit{Peer nodes}. Furthermore, Fabric supports configurable policies that define which peer nodes are required to participate in consensus for each blockchain transaction execution.

In the environment setup, our approach creates a Fabric network with a single orderer node running in solo mode to provide ordering services. One peer node is assigned to each membership. This is crucial because each membership must operate a peer node to validate SC executions and participate in consensus, thereby ensuring confidence in the blockchain execution results. By default, the consensus policy is set to the majority rule, requiring support from most members to validate and confirm the execution outcomes.

Additionally, Hyperledger Fabric employs an access control mechanism based on X.509 certificates, which serve two primary functions: 1) identity verification during SC invocation, and 2) TLS authentication for secure communication between nodes. To facilitate this, we establish a Certificate Authority (CA) for each membership to issue certificates for their nodes and users.

The entire on-chain environment setup in this work is developed based on Hyperledger Cello~\cite{cello}, which serves as a blockchain provision and operation system. Given that the new version of Hyperledger Cello is still under development and its functionalities are not yet fully matured\cite{Journal-software}, this paper has enhanced its capabilities for supporting Fabric CA and SC deployment. 

\textbf{For the off-chain layer}, in order to alleviate the problem of transporting large amounts of data within the blockchain, the IPFS~\cite{benet2014ipfs} serves as the central data repository, with data hashes utilized as descriptors in the blockchain to represent the actual data. By deploying an IPFS Node for each membership, a consortium-owned IPFS cluster is established, providing a robust and secure data source with high availability.

\subsubsection{\textbf{The connector linking on-chain and off-chain}}\label{subsection-connector}
\begin{figure}
    \centering
	\includegraphics[width=\linewidth]{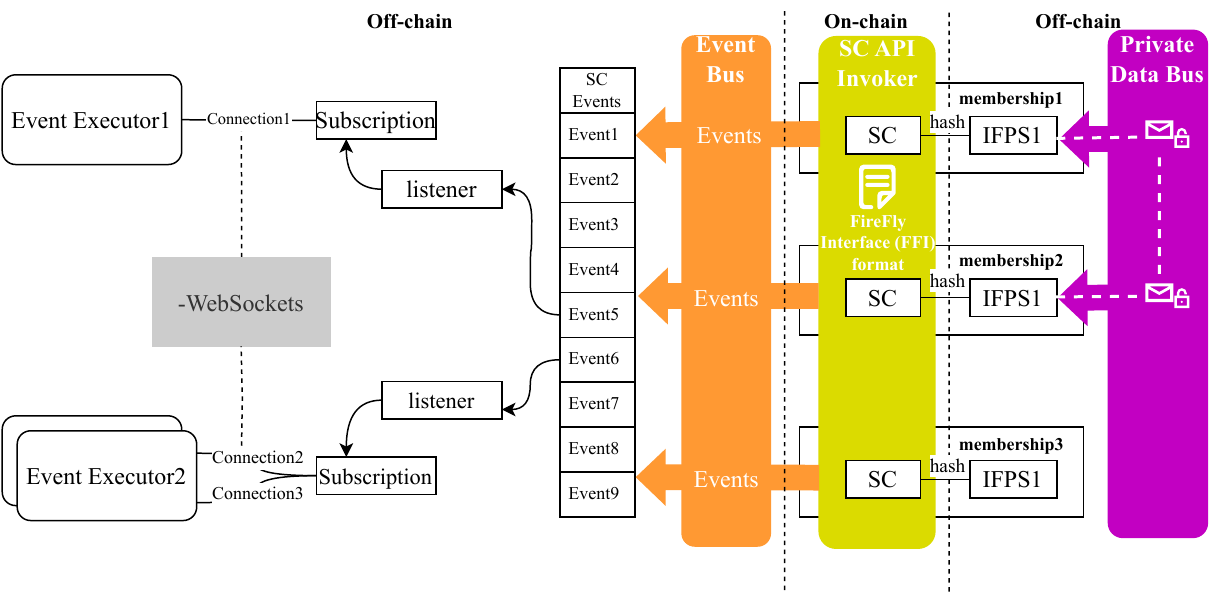}
	\caption{The connector linking on-chain and off-chain}
	\label{fig:event-bus}
\end{figure}

% \textbf{For the connector layer}, 
% Hyperledger FireFly~\cite{firefly} is an open source Supernode, a complete stack for enterprises to build and scale secure Web3 applications. \textit{\textbf{Firefly Organization}} registers client type certificates via the CA and links these to memberships. It also sets up connections with peers through initial configurations. \textbf{\textit{Firefly Node}} represent the organization's off-chain identities, with the Firefly Organization able to correspond to multiple Firefly Node. Additionally, this work utilizes Firefly's \textbf{\textit{Event Bus}} feature to support subscription to blockchain events, where the off-chain  \textbf{\textit{Event Executor}} performs specific actions upon detecting certain events. The  \textbf{\textit{SC API Invoker}} enables calling SCs through APIs, while the  \textbf{\textit{Private Data Exchange}} facilitates the off-chain transfer of sensitive data with a hash record stored on-chain. These details will be discussed in~\ref{subsection-connector}.
The Connector in our design functions as a bridge, integrating on-chain and off-chain resources to achieve specific functionalities that address challenges inherent in blockchain systems. These include delivering events to the external world, enabling convenient SC invocation, and handling data exchange efficiently and safely.

To achieve these objectives, we primarily use Hyperledger Firefly in our system by setting up a Firefly SuperNode for each membership. Each Firefly Supernode registers a client type certificate with a certificate Authority (CA), enabling it to interact with the blockchain as a user within the membership. These nodes are interconnected to form a \textit{Private Data Bus}, which facilitates off-chain data exchange and provides access to SCs on the blockchain through \textit{Event Bus} and the \textit{SC API Invoker}.

To facilitate the integration of blockchain-based SC applications into each organization’s existing systems, we use the \textit{FireFly Interface (FFI) format} to provide a common, blockchain-agnostic description of the SC, as shown in the yellow section of Fig.~\ref{fig:event-bus}. When the SC code is generated from BPMN choreography, an FFI file is also created. The feature of FireFly’s registration API then generates an HTTP API for this SC, complete with an OpenAPI Specification and Swagger UI. Once created, each organization will have its own API URL for the SC, enabling them to respectively invoke the SC.

In a decentralized system, each organization must run its own applications, integrating the shared state of the SC with its private data and core systems. Therefore, an \textit{Event Executor} is needed to continuously monitor blockchain events and sync these changes to the organization's private application state database. To enable this, we have integrated FireFly’s Event Bus feature, as shown in the orange part of Fig.~\ref{fig:event-bus}. The \textit{Event Bus} can capture all events in blocks, categorize them by topic based on specific fields in the event messages, and provide a subscription service to notify external systems of selected events. Each organization can register listeners and subscriptions to monitor events of interest and execute custom business logic within the Event Executor.

In multi-party collaboration scenarios, the ability to exchange private data is essential. Private data interaction is the primary communication mode for many organizations today, where one party sends data to another through a secure, mutually agreed-upon channel. In Firefly, the \textit{Private Data Bus} component facilitates this private data exchange, as illustrated in the purple part of Fig.~\ref{fig:event-bus}. During the \textit{Instance creation phase}, membership is associated with participants in BPMN Choreography. In the \textit{Collaboration and decision execution phase}, the \textit{Private Data Bus}’s private send function is triggered to enable off-chain private message exchanges between the Initiator and Recipient in BPMN, while also storing proof on-chain. This approach ensures message confidentiality while maintaining transparency in the collaborative process.

\subsubsection{\textbf{Oracle connects reality with blockchain}}\label{subsubsection-oracle}

To enable SCs to access off-chain data, our method integrates an \textit{Oracle}, which links the blockchain to external data sources. Part of the Oracle operates on-chain to handle requests, while another part works off-chain to collect and validate data. While typically used to retrieve off-chain data, we also consider a reverse oracle mode that supplies on-chain data externally.

Data flow through Oracles is generally classified into four patterns, based on data direction and request initiation~\cite{oracle_pattern}: \textit{Inbound} and \textit{Outbound Oracles}, which indicate the data's movement to or from the blockchain, and \textit{Pull-based} and \textit{Push-based} request models, which specify how requests are triggered.
\begin{itemize}
    \item \textbf{Pull-based inbound Oracle}: The SC requests data from the oracle, which fetches it from an off-chain source.
    \item \textbf{Push-based inbound Oracle}: Oracle watches for any sort of changes in a particular off-chain data source. 
    \item \textbf{Pull-based outbound Oracle}: When an off-chain resource needs to query data from the blockchain, it requests the data from the on-chain data source. 
    \item \textbf{Push-based outbound Oracle}: The SC monitors the blockchain for changes and informs the off-chain resource.
\end{itemize}

In our implementation, we need to store and retrieve data from an IPFS cluster and provide off-chain data access. Thus, we implement \textit{Push-based Outbound Oracle}, \textit{Pull-based Inbound Oracle}, and \textit{Pull-based Outbound Oracle} to support our system.

\begin{figure}
    \centering
	\includegraphics[width=\linewidth]{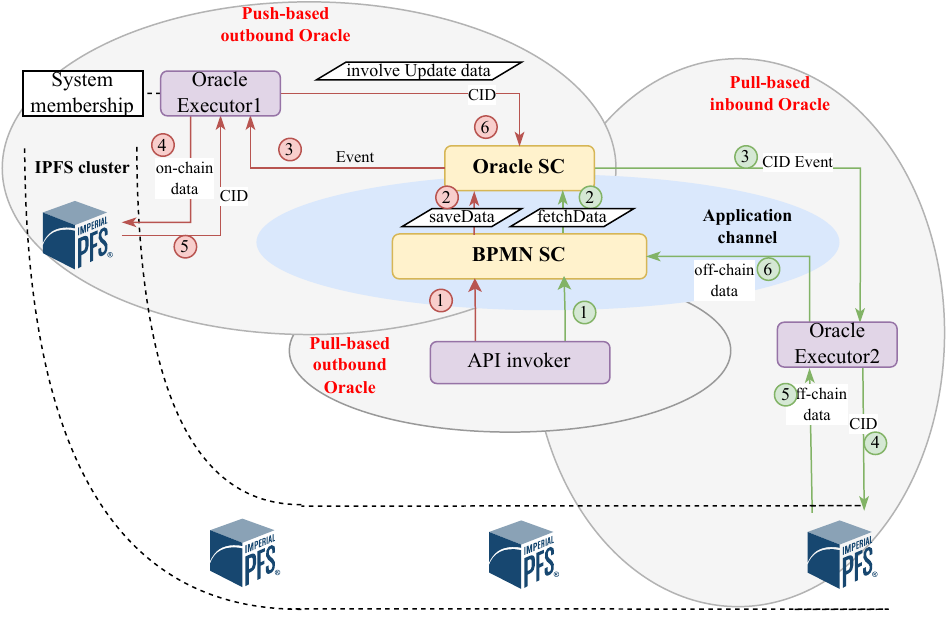}
	\caption{The implementation architecture of Oracle in this paper}
	\label{fig:Oracle}
\end{figure}

Fig.~\ref{fig:Oracle} illustrates our Oracle mechanism, which operates through a collaborative framework between the Oracle SC and Oracle Executors, enabling seamless data storage and retrieval on IPFS. The Oracle SC processes requests from other BPMN SCs on-chain and broadcasts events to off-chain Oracle Executors, which handle data retrieval or storage depending on the event type. To maintain data integrity, only entities with system membership are authorized to perform write operations, while data queries remain accessible to all SCs.

The Oracle SC is designed as a independent SC, adaptable across different SCs for data retrieval and storage in IPFS. It provides two primary methods: \textit{saveData} and \textit{fetchData}. The saveData method receives data and a key, emits an event with the data content, and waits for the Executor to store the content and upload the corresponding CID for record-keeping. The fetchData method accepts a query with a record ID and callback method as parameters, emits an event for the Executor to retrieve data from IPFS, and invokes the callback with the requested data.

As shown in Fig.~\ref{fig:Oracle}, our implementation of these Oracle patterns enables essential system functions. The \textit{Push-based Outbound Oracle} (left) uses the \textit{saveData} method to transfer data from on-chain to off-chain. The \textit{Pull-based Inbound Oracle} (right) retrieves off-chain data on-chain via the \textit{fetchData} method. The \textit{Pull-based Outbound Oracle} (center) provides external users with on-chain data access through an API.

\subsection{Smart Contract Generate Method}\label{sect:SC-ganerate}
To support the execution of the BPMN and DMN hybrid model proposed in Sect.~\ref{subsection:model-method}, a methodology is presented for mapping the business model into SC. This methodology comprises three main parts: 1) Implementing collaborative decisions execution utilizing a DMN decision engine to support complex and configurable decision logic; 2) Employing ABAC methods; and 3) Collaborative business processes execution based on a state machine approach. 

\subsubsection{\textbf{Decison service (DS) to support decision execution}}
According to the design in Section~\ref{subsection:model-method}, this paper implements a DS for SC (hereafter referred to as DMN SC). Given the input data defined in the BPMN choreography BRT, it calculates and provides outputs based on the specified DMN decision models. In addition to executing decision outcomes, the DMN SC must also record the process data of each decision (inputs, outputs, and the unique ID of the DMN) and store it on the blockchain. This serves as the universally recognized decision outcome among organizations, facilitating auditors in tracing the origins for future audits.

Since many traditional DMN engines (such as Drools \cite{proctor2012drools}) support the execution of DMN which are written in various languages, this paper does not develop a new SC to parse and execute DMN from scratch. Moreover, as the Fabric platform officially supports SC in Go, Java, and JavaScript, this paper integrates a lightweight, open-source DMN engine written in Java: the Camunda DMN engine.

A DMN file may contain highly complex decision logic, making it too large to store directly on the blockchain. Therefore, in our method, the DMN content is stored off-chain, with retrieval and upload occurring only during execution. During the instance creation phase, users must bind a DMN to each BRT, which is saved to off-chain IPFS using a Push-based outbound Oracle through the \textit{saveData} method, resulting in a CID recorded within the Oracle SC, as shown in Fig.~\ref{fig:Oracle}. Additionally, a hash digest of the DMN content is calculated and saved on-chain to ensure data integrity.

In the Execution Phase, when a BRT element becomes active, the actual DMN content and input data are provided. Also illustrated by Fig.~\ref{fig:Oracle}, by invoking the \textit{fetchData} method, a Pull-based inbound Oracle is activated to retrieve the DMN content using the CID recorded during the instance creation phase. The hash digest is then recalculated to verify consistency with the previous hash, ensuring data integrity. Once verified, the DMN content and input data are passed to the DMN SC, where the decision logic is executed, producing the desired decision result.

\subsubsection{\textbf{Collaborative Identity Management based on Blockchain ABAC}}

In our method, collaboration relies on the coordinated actions of all participating organizations, each of which must define and take responsibility for its specific role and related tasks. Only the designated organization may execute the tasks associated with its role. For instance, in the Supply Chain scenario illustrated in Figure~\ref{fig:bpmn-running-example}, the organization in the Middleman role is solely authorized to perform tasks like "Supply Order Forwarding" and "Transport Order Forwarding". Additionally, custom restrictions, such as requiring the task invoker to have at least ten years of experience, may apply. A mechanism is therefore needed to enforce these requirements.

The blockchain platform used in our work, Hyperledger Fabric, provides access control to limit Peer Node access and permissions to modify chaincode across the network by requiring an X.509 certificate to verify the Membership Service Provider (MSP). However, it lacks a participant-level access control mechanism. To address this, we introduce an ABAC method to regulate access to SCs and their instances.

ABAC is a policy-based access control mechanism that defines access based on user, resource, and contextual attributes. Our method uses information from the X.509 certificate, which includes MSP and embedded attributes from issuance. During instance creation, role-specific restrictions in the BPMN such as organizational assignment and requirements like minimum age or experience must be defined. Before modifying the blockchain state in Execution, each task invocation undergoes access control verification, as shown in Algorithm 1 in Appendix, ensuring authorization for each task.

\subsubsection{\textbf{State Machine Based Workflow Control}}\label{sub-sect:state-machine}

As mentioned, organizations collaborate by executing tasks within the same SC, but what content does the SC provide? Our approach identifies three primary objectives to fulfill through the SC generation method: 1) providing methods to ensure that processes advance in the correct order as defined in the BPMN choreography, 2) enabling message exchange and validation between two organizations, and 3) supporting the execution of BRTs with off-chain DMN content.

\begin{figure}[htbp]
    \centering
	\includegraphics[width=\linewidth]{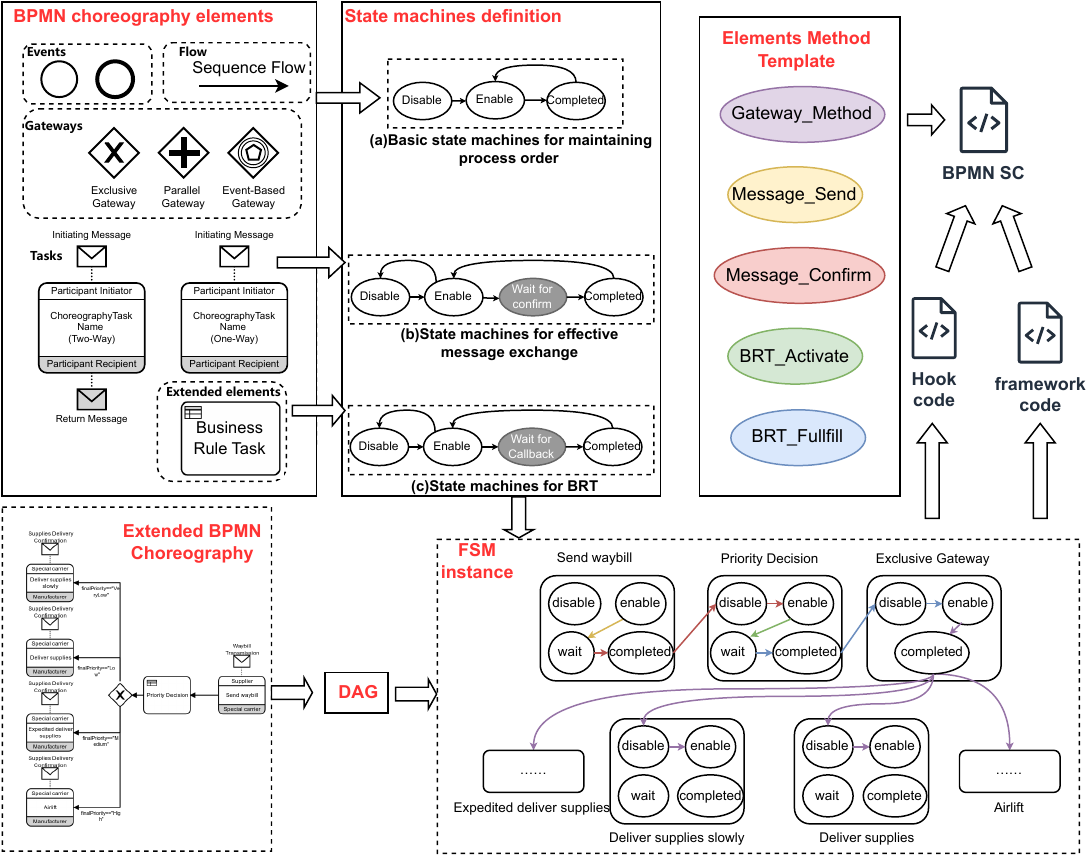}
	\caption{Integrated BPMN-DMN models to SC transformation method}
	\label{fig:fsm-and-methods}
\end{figure}

To achieve these objectives, our approach proposes a method for transforming the integrated BPMN-DMN models into executable SC. As shown in Figure~\ref{fig:fsm-and-methods}, the extended BPMN choreography is parsed into a Directed Acyclic Graph (DAG), where each element is converted into a finite state machine (FSM) with three or four states. These FSMs are then organized according to the choreography’s topology. Using the DAG, we analyze metadata, including Hook Code and Frame Code, which are combined with element templates to produce the final SC.
\textbf{To maintain the sequence of process}, our approach manages the extended BPMN choreography through states within elements and workflows between elements. In our method, each element in the model is constructed as an FSM with at least three states: Disabled, Enabled, and Completed. Additional states are introduced as needed, which are discussed further below. As illustrated in Algorithm. 3 in Appendix, each method template includes a state check to ensure the current element is in the Enabled state, preventing out-of-order execution.

At a higher level, these elements together form a global FSM, where Sequence Flows and Gateways regulate the order of individual FSMs. Generally, Sequence Flows dictate the primary workflow sequence, while Gateways manage more complex flow patterns.

\begin{itemize} \item \textit{Parallel Gateways} serve dual roles: splitting and merging. In the splitting role, all outgoing sequence flows are enabled when the element is completed. As a merger, the gateway enables the outgoing sequence flow once all incoming flows are completed. \item \textit{Exclusive Gateways} also function in two roles. As a splitter, it enables sequence flows based on conditions; as a merger, it activates the next sequence flow once any incoming flow is completed. In our method, we use conditional expressions in sequence flows to represent complex decision policies within Business Rule Tasks (BRTs). \item \textit{Event-based Gateways} provide unique functionality by enabling sequence flows based on triggered events. However, selecting one sequence flow deactivates the others. \end{itemize}

\textbf{Effective message exchange} is achieved through a process involving data credential upload and message content verification. In the first phase, the message sender pre-stores the message in the \textit{Private Data Bus}, generates a message hash as proof, and uploads this hash to notify the receiver. In the second phase, the receiver retrieves the message content using the hash, verifies it, and, if valid, confirms the message. To support this process, the message element includes four states instead of three, with an additional “Wait for Confirm” state indicating that the message is pending confirmation from the receiver. These two phases are implemented with two template methods, \textit{Message} and \textit{MessageConfirm}.

\textbf{The process of fetching DMN content} is handled similarly through a two-part process, corresponding to two template methods shown in Algorithm. 2 in Appendix: \textit{BusinessRuleTask} and \textit{BusinessRuleTaskCallback}. The former triggers an event in the Event Bus, activating the Pull-based Inbound Oracle to retrieve the DMN content. Subsequently, the Oracle Executor invokes \textit{BusinessRuleTaskCallback}. Within the callback, the DMN content is verified against the content provided during instance creation. Decision-making parameters are then extracted from the instance context and input into the DMNEngine SC to reach a decision. To support this process, an additional “Wait for Callback” state is added to the BRT element.

\begin{table*}
\caption{Qualitative Comparison of Method and Related Work Features}
\centering
\label{table:comparison}
\resizebox{\textwidth}{!}{% 适应页面宽度
\begin{tblr}{
  width = \linewidth,
  colspec = {Q[120]Q[65]Q[65]Q[50]Q[50]Q[50]Q[40]Q[80]Q[63]Q[54]Q[62]Q[150]},
  cells = {c},
  cell{1}{1} = {r=2}{},
  cell{1}{2} = {r=2}{},
  cell{1}{3} = {r=2}{},
  cell{1}{4} = {c=4}{0.272\linewidth},
  cell{1}{8} = {r=2}{},
  cell{1}{9} = {r=2}{},
  cell{1}{10} = {r=2}{},
  cell{1}{11} = {r=2}{},
  cell{1}{12} = {r=2}{},
  hline{1,12} = {-}{0.08em},
  hline{2} = {4-7}{l},
  hline{3} = {-}{0.05em},
}
{MED\\ Research}                                                       & {Blockchain\\ Platform}          & {BPMN\\Model}      & Oracle                    &                           &                           &                           & {Data\\ Privacy}      & {External\\ Integration} & {Complex\\ Decision\\ Making} & {Access\\ Control} & Feature \\&                                  &                       & {Inbound\\ Push}          & {Inbound\\ Pull}          & {Outbound\\ Push}         & {Outbound\\ pull}         &                       &                          &                               &                    &          \\
BlockCollab                                                            & {Fabric}            & {Choreography}  & \xmark & \cmark & \cmark & \cmark & {Off-Chain\\Data Bus} & Restful API              & Gateway, DMN                           & ABAC               & Integreated BaaS                                     \\
CaterPillar~\cite{Caterpillar2017,Caterpillar2019}    & Ethereum                         & {Collaboration} & \xmark & \cmark & \cmark & \cmark & N/A                   & Restful API              & Gateway                       & {Account\\Based}   & SubProcess                                           \\
Lorikeet~\cite{Lorikeet,LorikeetJournal}              & Ethereum                         & {Collaboration} & \xmark & \cmark & \cmark & \cmark & N/A                   & N/A                      & Gateway                       & {Account\\Based}   & Asset Management                                     \\
UBPM~\cite{weber_choreography}                       & Ethereum                         & {Choreography}  & \xmark & \cmark & \cmark & \cmark & {Encrypting\\Payload} & N/A                      & Gateway                       & {Account\\Based}   &                                                      \\
ChorChain~\cite{chor-chain,chor-chain2}               & Ethereum                         & {Choreography}  & \xmark & \xmark & \xmark & \cmark & N/A                   & N/A                      & Gateway                       & {Account\\Based}   & Auditing Feature                                     \\
FlexChain~\cite{FlexChain1,FlexChain2}                & Ethereum                         & {Choreography}  & \xmark & \xmark & \xmark & \cmark & N/A                   & N/A                      & Gateway                       & {Account\\Based}   & {Drools-based state checking} \\
MultiChain~\cite{corradini2021model_multiBlockchain} & {Ethereum,\\Fabric} & {Choreography}  & \xmark & \xmark & \xmark & \cmark & N/A                   & N/A                      & Gateway                       & ABAC               & {Support Two Blockchain}                   \\
ECBS~\cite{renmin-university}                         & {Fabric}            & {Collaboration} & \xmark & \xmark & \xmark & \cmark & N/A                   & N/A                      & Gateway                       & MSP         & {Process Engine instead of SC Generation}    \\
IBC~\cite{shenIcws}                                   & {Fabric}            & {Choreography}  & \xmark & \xmark & \xmark & \cmark & Off-Chain Data Bus                   & Restful API                  & Gateway                       & MSP         & Integrated BaaS                                      
\end{tblr}}
\end{table*}

We then integrate these three design components to generate the SC through a structured two-pass iteration. As shown in Algorithm. 2 in Appendix, the model is first parsed into a DAG and processed with the \textit{GenerateHooks} function to analyze the topology of each element. This process produces "HookCode"—code segments embedded within template methods at predefined positions, enabling customized transitions between states. In the second pass, the model is parsed again to assemble methods for each element using method templates. Finally, the template code and corresponding HookCode are then combined to produce the executable SC.

\section{Experiment}

\subsection{Qualitative Comparison}\label{sect:exper-function-comparison}
To highlight the differences between our method and related work, we reviewed these studies and selected specific functional features as comparison criteria, allowing us to evaluate our approach and others within the same framework.

After synthesizing these works, we identify a set of common criteria for critique. \textit{Blockchain Platform} denotes the specific blockchain technology on which each work is based or compatible. \textit{BPMN Model} indicates the BPMN diagram type used to model collaborative processes. \textit{Oracle} refers to the data flow patterns as metioned in Sect.~\ref{subsubsection-oracle} the work employs. \textit{Data Privacy} assesses whether the work incorporates methods for secure data exchange between participants. \textit{External Integration} evaluates the ease with which external systems can integrate with the work. \textit{Complex Decision Making} considers the method used to model sophisticated decision-making behaviors within processes. \textit{Access Control} identifies the type of access control mechanisms implemented. Finally, \textit{Feature} highlights any distinctive features that set the work apart from others.

The synthesis results, illustrated in Table.~\ref{table:comparison}, reveal that most existing works primarily target permissionless blockchains, whereas our approach is designed for permissioned blockchain environments. Compared to these works, our approach offers enhanced capabilities in Data Privacy and External Integration. Additionally, building upon previous work in IBC, we integrate advanced Complex Decision-Making methods and a robust Access Control Mechanism. Although other works may feature unique elements such as SubProcess and Asset Management, these were not within the primary scope of our study.

\begin{table*}
    \centering
    \caption{The verification of the correctness of SC}
    \label{table:noise_result}
    \resizebox{\textwidth}{!}{% 适应页面宽度
    \begin{threeparttable}
    \begin{tabular}{@{}lcccccccccc@{}}
        \toprule
        \textbf{Scenarios} & \textbf{\#Generate paths} & \textbf{\#Basic Path} & \textbf{\#Conforming} & \textbf{\#Not-conforming} & \textbf{\#Tasks} & \textbf{\#Message} & \textbf{\#Gateways} & \textbf{\#BRTs} & \textbf{Accuracy} \\ 
        \midrule
        Hotel booking~\cite{chor-chain}            & 1313            & 10                     & 22                  & 1305                  & 9                     & 13                  & 6                    & 1                   & 100\%              \\
        Customer~\cite{shenIcws}        & 1285            & 9                      & 18                 & 1267                  & 8                     & 13                  & 6                    & 2                   & 100\%              \\
        The article example(Fig.~\ref{fig:bpmn-running-example})     & 492             & 4                      & 11                 & 481                   & 13                    & 13                  & 4                    & 1                   & 100\%              \\
        Supply Chain~\cite{weber_choreography}    & 332             & 2                      & 4                  & 328                   & 11                    & 11                  & 3                    & 1                   & 100\%              \\
        Blood Analysis~\cite{chor-chain}    & 472             & 3                      & 12                 & 460                   & 6                     & 6                   & 3                    & 1                   & 100\%              \\
        Amazon SLA\tnote{1}         & 637             & 4                      & 8                  & 629                   & 8                     & 8                   & 3                    & 1                   & 100\%              \\
        Pizza Order\tnote{2}            & 409             & 3                      & 6                  & 403                   & 8                     & 8                   & 4                    & 1                   & 100\%              \\
        Rental Claim~\cite{Rental-Claim}& 334             & 4                      & 8                  & 326                   & 8                     & 8                   & 3                    & 1                   & 100\%              \\
        Purchase~\cite{Purchase}         & 459             & 4                      & 8                  & 451                   & 7                     & 7                   & 3                    & 1                   & 100\%              \\
        Manufactory \tnote{3}    & 503             & 3                      & 6                  & 497                   & 6                     & 10                  & 2                    & 1                   & 100\%              \\
        Management system~\cite{management-system}       & 684             & 4                      & 8                  & 676                   & 6                     & 6                   & 3                    & 1                   & 100\%              \\ 
        \bottomrule
    \end{tabular}
    \begin{tablenotes}
        \item [1] https://aws.amazon.com/cn/compute/sla
        \item[2] https://camunda.com/blog/2021/01/chor-js-an-editor-for-bpmn-choreography-diagrams/
        \item [3] https://www.slideserve.com/osias/business-process-modelling-using-bpmn-part-ii
    \end{tablenotes}
    \end{threeparttable}
    }
\end{table*}

\subsection{Applicability of the Modeling Method and Correctness of the Generated SC Code}
To evaluate the effectiveness and applicability of the proposed modeling method, we selected 11 scenarios from publicly available real-world cases and published studies (as shown in the "Scenarios" column in Table~\ref{table:noise_result}), utilizing the method described in Sect.~\ref{subsection:model-method} to model business processes and decisions for multi-party collaboration. These cases span various domains, demonstrating the applicability and effectiveness of the proposed modeling approach in diverse collaborative scenarios.

As outlined in Sect.~\ref{sub-sect:state-machine}, the generated SCs are specifically designed to ensure the correct execution of process sequences. In this section, we conduct experiments to evaluate whether the generated SC can maintain the sequential execution required for the choreography process. Following the method described in paper~\cite{weber_choreography}, we generate both \textit{\textbf{Conforming}} and \textit{\textbf{Not-conforming}} paths based on the 11 scenarios. We then execute these paths through the SC to verify its ability to handle both compliant and non-compliant sequences.

The experimental procedure is as follows: Firstly, for the choreography diagram of each scenario, the number of \textit{Basic path} is analyzed based on the DAG and the number of gateways (If there is a loop, we only list the normal path and the path that loops once as two basic path; if there are elements between two parallel gateways, we also list only one sequential path as a basic path). Secondly, based on the LLM, the basic paths and BPMN choreography diagrams are sent to the LLM to assist in generating message parameters corresponding to different basic paths. Thirdly, for each basic path, the following operations are randomly applied to modify these paths and generate a larger set of test paths: (i) add an element on the path to a different position in the original path, (ii) remove an element from the original path, and (iii) swap two elements in the original path. Finally, each path is executed on the third-party collaboration platform, and the results are examined and summarized in Table.~\ref{table:noise_result}.

After conducting manual verification, we found that all results aligned with the expected outcomes, achieving a 100\% accuracy rate, which demonstrates that the SC code successfully maintains the intended behavior.

\subsection{The Usability of The Third-Party Collaboration Platform in the Execution Phase}

This section’s experiment primarily aims to evaluate the usability of the third-Party collaboration platform in the execution phase, where different elements follow distinct processes within the system, categorized into three types as described below:

\begin{itemize} 
\item \textit{Event/Gateway}: The simplest type, where an API Invoker calls the SC method and waits for BPMN SC to reach consensus.
\item \textit{Message}: More complex, involving interaction with the Private Data Bus. A message is sent to the Private Data Bus, which saves the content in IPFS, generates a CID, and creates a message item linked to it. The message ID is returned and used to invoke BPMN SC, following the same consensus process as before.
\item \textit{BRT}: The most complex type, which begins by sending an event to Firefly’s event bus. Firefly forwards it to the Oracle Executor, triggering it to retrieve DMN content from IPFS via CID, invoke a callback with the content, and then proceed to consensus.
\end{itemize}

Since the first type is also included in the other two, we will only set up experiments for the latter two.

To illustrate the process, we use sequence diagrams and timelines. The sequence diagram outlines message flows between components, while the \textit{Timeline Steps} breaks the process into multiple sections, each representing a key system component or SC. This timeline records the average time taken for each segment across 100 test runs.

\begin{figure}
    \centering
	\includegraphics[width=\linewidth]{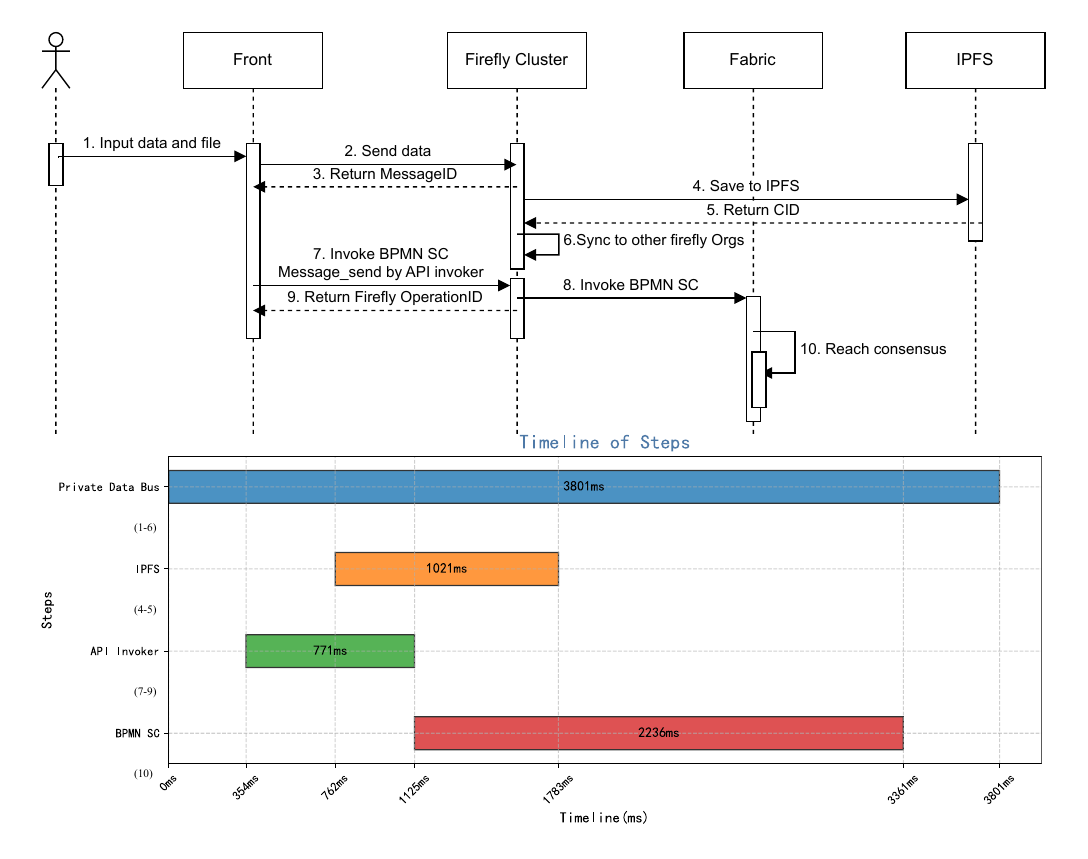}
	\caption{Sequence diagram and timeline of \textit{Message} execution}
    \label{fig:experiment-performance1}
\end{figure}
As shown in Figure.~\ref{fig:experiment-performance1}, the entire process of \textit{Message} takes approximately 3.8 seconds. The Private Data Bus handles flows 1–6, managing data storage on IPFS and synchronizing data across organizations. The Fabric network requires about 2236 ms for SC invocation and consensus, largely due to the 2-second block generation interval of Fabric. From the user’s perspective, the process appears complete once SC consensus is achieved, while IPFS storage and data synchronization continue in the background. This results in an average completion time of around 3361 ms.

\begin{figure}
    \centering
	\includegraphics[width=\linewidth]{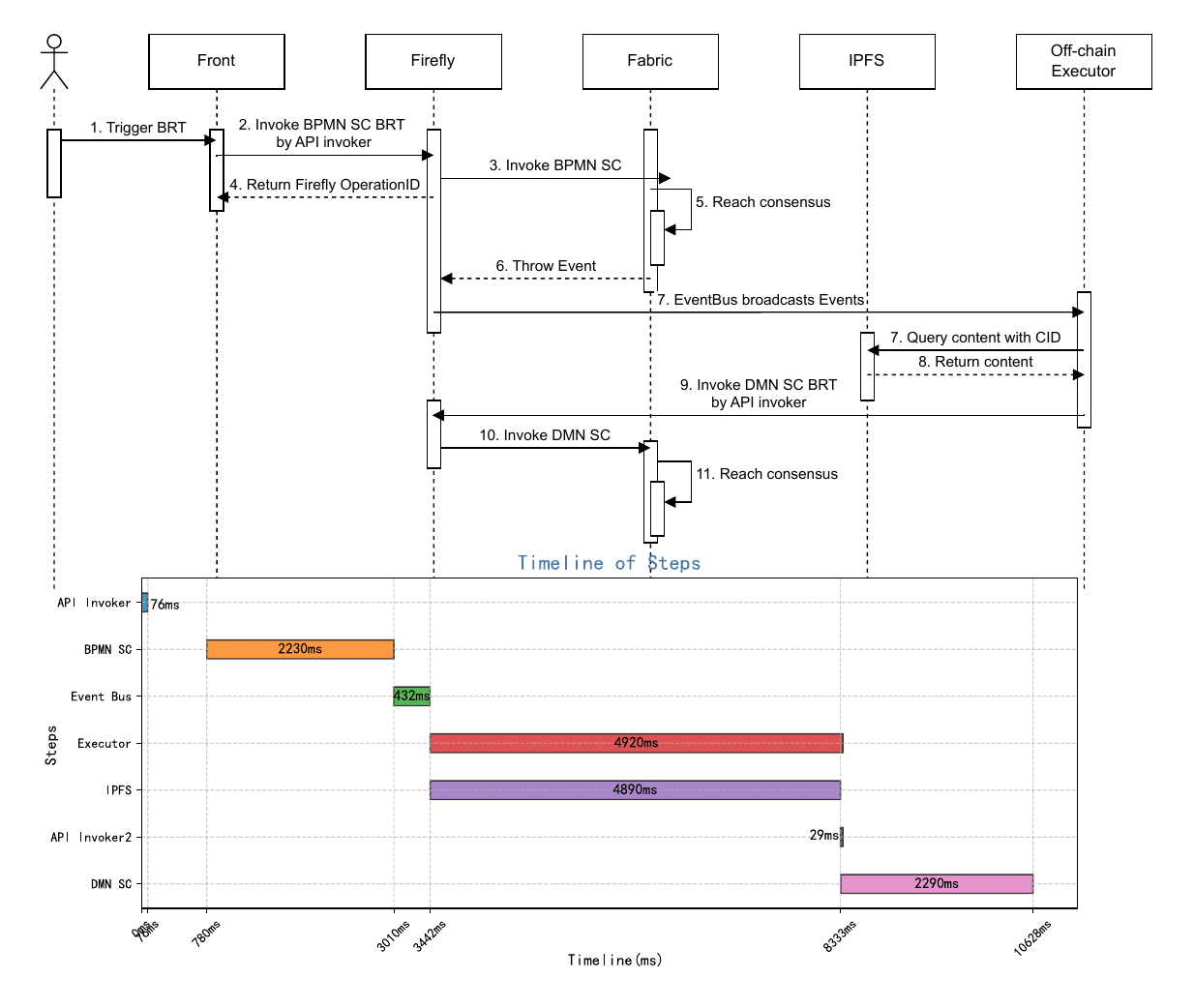}
    \caption{Sequence diagram and timeline of \textit{BRT} execution}
    \label{fig:experiment-performance2}
\end{figure}

Figure.~\ref{fig:experiment-performance2} illustrates the \textit{BRT} process, which is more complex than \textit{Message}. \textit{BRT} requires two SC invocations, each taking approximately 2 seconds due to block intervals. Additionally, IPFS read times are close to 5 seconds, creating a primary bottleneck. Users must wait for DMN SC consensus before they can see updated states, resulting in a total duration of 10628 ms. Since \textit{BRT} is relatively infrequent, this delay is generally considered acceptable.

\section{Related Works}\label{sect:related work}

\subsection{Leveraging Blockchain for Trustworthy Multi-Party Collaborative Processes Execution}\label{sec:sub:related-bpmn-blockchain}
Most research focuses on using specific languages to model inter-organizational collaboration processes, particularly the interaction of activities across organizational boundaries. Many studies use BPMN for modeling, which includes collaboration and choreography diagrams. This paper chooses the choreography diagram, as it is especially suitable for scenarios involving multiple parties~\cite{OMG2013BPMN}. Mendling, Jan et al.~\cite{weber_survey} suggest exploring how blockchain can transform particular processes and foster collaboration with external stakeholders. This technology has the potential to rejuvenate the entire field of choreography.

In order to support the trustworthy execution of collaborative processes, previous research has introduced blockchain technology, which is considered as a promising technical solution for revitalizing the research area~\cite{renmin-university}. Previous research has mostly used the permissionless blockchain, focusing only on the generation of solidity SCs for supporting multi-party collaboration. This paper adopts the permissioned blockchain, focusing on privacy and efficiency issues in multi-party collaboration processes. However, this introduces the complexity brought about by the deployment of blockchain infrastructure.

The integration of BPMN and blockchain has been proposed in numerous research works to support multi-party collaboration processes. F. Corradini et al. proposed the ChorChain framework~\cite{chor-chain,chor-chain2}, which uses a model-driven approach to generate Solidity SCs from BPMN choreography diagrams, supporting a trustworthy and auditable blockchain orchestration system and covering the entire lifecycle from modeling to distributed execution and auditing. They proposed the Multi-Chain framework~\cite{corradini2021model_multiBlockchain}, which uses a model-driven approach to generate Fabric SCs and Ethereum Solidity SCs from BPMN choreography, meeting different privacy requirements.

Weber, Ingo et al. proposed the first open-source execution engine named Caterpillar~\cite{Caterpillar2017,Caterpillar2019} for BPMN collaboration diagrams on Ethereum, combining the convenience of Business Process Management Systems (BPMS) with the decentralized features of blockchain platforms. \cite{weber_choreography} is the method proposed by Weber et al. for executing BPMN choreography on Ethereum, which addresses trust issues in cross-organizational collaboration without the need for an authoritative body. \cite{Blockchain-Support} summarizes two tools, one of which is a tool called Lorikeet~\cite{Lorikeet,LorikeetJournal}, integrating the registry editor Regerator and implementing BPMN translation algorithms from both \cite{weber_choreography} and \cite{Optimized}.

Nakamura, Hiroaki et al. \cite{Inter-organizational-by-blockchain} utilize elements from BPMN process models to represent shared processes between organizations. These models are subsequently converted into statecharts to facilitate the development of web applications and the creation of SCs.

However, these research merely support the generation of SC code for permissionless blockchains, with most focusing on Ethereum’s Solidity SC. Meanwhile, the complexity of the underlying infrastructure introduced by permissioned blockchains is ignored.

\subsection{Integrating Decisions with Processes Supported by Blockchain}\label{sec:sub:related-dmn}

% The previous section (Sect.~\ref{sec:sub:related-bpmn-blockchain}) has analyzed the business processes using blockchain in multi-party collaboration. 

In business process management, decisions are now more deeply embedded within processes~\cite{BPMN-dmn-separation}. This integration enhances organizational competitiveness, allowing decisions to be analyzed, implemented, and reused across various processes to optimize business outcomes~\cite{Real-world-DMN}. To separate decisions from processes and prevent BPMN from becoming overly redundant and complex, the DMN was developed by the Object Management Group for modeling decisions at different levels of detail. As mentioned in~\cite{MDE4BBIS}, it would be interesting to utilize blockchain technology with DMN to represent the execution of specific decision activities.

Stephan Haarmann et al. attempted to execute decision problems on the Ethereum blockchain by converting S-FEEL expressions from DMN into Solidity code to achieve immutable decision logic~\cite{DMN-execution-ethereum}. Subsequently, they explored privacy issues in executing collaborative decisions on blockchain technology and proposed a new method to support decision-making without exposing sensitive data~\cite{execution-decision}. Stephan Haarmann further improved his previous work in research~\cite{Executing-DMN}, by enabling privacy-preserving decision execution and semi-automated conflict resolution.

Flavio Corradini et al. introduced the FlexChain framework~\cite{FlexChain1,FlexChain2}, which decouples logic from execution states in BPMN, utilizing an on-chain/off-chain architecture where the Drools rule engine executes BPMN decision logic off-chain.

In summary, prior studies have yet to realize the direct execution of DMN within workflows, and there has been limited exploration of implementations on permissioned blockchains. Furthermore, off-chain execution of decision components may face tampering risks and consistency issues.

\section{Conclusion and Discussion}
In conclusion, this paper presents a novel model-driven approach that combines BPMN choreography with DMN to model collaborative business processes and decisions in a blockchain setting. This methodology supports the automatic generation of SC code for blockchain execution. Furthermore, we introduce an innovative hybrid on-chain and off-chain execution environment, enhancing the integration of blockchain with real-world systems. In the end, the functionalities of the paper were developed onto a third-party collaboration platform, which was then extensively validated through experiments based on this platform.

However, there are certain shortcomings in the current work. Firstly, different organizations may have heterogeneous blockchain environments, and there is a need for collaborative scenarios based on permissionless blockchain(e.g. Ethereum). Therefore, future work should explore how to integrate this platform with multiple blockchain environments and support cross-chain operations between different blockchains, while also generating SC code tailored for diverse blockchain environments. Secondly, the current blockchain-based collaboration modeling is not sufficiently comprehensive. It could further incorporate blockchain asset management and enhance the design of Oracles.

% or


\section*{Acknowledgment}

The research presented in this paper has been partially supported by the National Science Foundation of China (62372140), Key Research and Development Program of Heilongjiang Providence (2022ZX01A28), Postdoctoral Fellowship Program of CPSF under Grant Number GZC20242204 and Heilongjiang Posterdoctoral Funding (LBH-Z23161).

\ifCLASSOPTIONcaptionsoff
  \newpage
\fi



%\bibliographystyle{IEEEtran}

% References section
\footnotesize{
\bibliographystyle{IEEEtran}
\bibliography{ref}

\begin{thebibliography}{10}
\providecommand{\url}[1]{#1}
\csname url@samestyle\endcsname
\providecommand{\newblock}{\relax}
\providecommand{\bibinfo}[2]{#2}
\providecommand{\BIBentrySTDinterwordspacing}{\spaceskip=0pt\relax}
\providecommand{\BIBentryALTinterwordstretchfactor}{4}
\providecommand{\BIBentryALTinterwordspacing}{\spaceskip=\fontdimen2\font plus
\BIBentryALTinterwordstretchfactor\fontdimen3\font minus \fontdimen4\font\relax}
\providecommand{\BIBforeignlanguage}[2]{{%
\expandafter\ifx\csname l@#1\endcsname\relax
\typeout{** WARNING: IEEEtran.bst: No hyphenation pattern has been}%
\typeout{** loaded for the language `#1'. Using the pattern for}%
\typeout{** the default language instead.}%
\else
\language=\csname l@#1\endcsname
\fi
#2}}
\providecommand{\BIBdecl}{\relax}
\BIBdecl

\bibitem{OMGDMN}
\BIBentryALTinterwordspacing
{Object Management Group}, ``Decision model and notation(dmn), version 1.5,'' Object Management Group, Dec. 2024, accessed: 2024-09-30. [Online]. Available: \url{https://www.omg.org/spec/DMN}
\BIBentrySTDinterwordspacing

\bibitem{haarmann2019executing_dmn}
S.~Haarmann, K.~Batoulis, A.~Nikaj, and M.~Weske, ``Executing collaborative decisions confidentially on blockchains,'' in \emph{Business Process Management: Blockchain and Central and Eastern Europe Forum: BPM 2019 Blockchain and CEE Forum, Vienna, Austria, September 1--6, 2019, Proceedings 17}.\hskip 1em plus 0.5em minus 0.4em\relax Springer, 2019, pp. 119--135.

\bibitem{lauster2020literature_survey_BCandBPMN}
C.~Lauster, P.~Klinger, N.~Schwab, and F.~Bodendorf, ``Literature review linking blockchain and business process management,'' in \emph{Proc. 15th Int. Conf. Wirtschaftsinformatik}, 2020, pp. 1802--1817.

\bibitem{weber_choreography}
I.~Weber, X.~Xu, R.~Riveret, G.~Governatori, A.~Ponomarev, and J.~Mendling, ``Untrusted business process monitoring and execution using blockchain,'' in \emph{Business Process Management}, M.~La~Rosa, P.~Loos, and O.~Pastor, Eds.\hskip 1em plus 0.5em minus 0.4em\relax Cham: Springer International Publishing, 2016, pp. 329--347.

\bibitem{inter_privateProcess}
H.~Nakamura, K.~Miyamoto, and M.~Kudo, ``Inter-organizational business processes managed by blockchain,'' in \emph{Web Information Systems Engineering--WISE 2018: 19th International Conference, Dubai, United Arab Emirates, November 12-15, 2018, Proceedings, Part I 19}.\hskip 1em plus 0.5em minus 0.4em\relax Springer, 2018, pp. 3--17.

\bibitem{OMG2013BPMN}
\BIBentryALTinterwordspacing
{Object Management Group}, ``Business process model and notation (bpmn), version 2.0.2,'' Object Management Group, Dec. 2014, accessed: 2024-03-10. [Online]. Available: \url{http://www.omg.org/spec/BPMN}
\BIBentrySTDinterwordspacing

\bibitem{haarmann2018dmn_execution}
S.~Haarmann, K.~Batoulis, A.~Nikaj, and M.~Weske, ``Dmn decision execution on the ethereum blockchain,'' in \emph{Advanced Information Systems Engineering: 30th International Conference, CAiSE 2018, Tallinn, Estonia, June 11-15, 2018, Proceedings 30}.\hskip 1em plus 0.5em minus 0.4em\relax Springer, 2018, pp. 327--341.

\bibitem{fernandez2013camunda}
A.~Fernandez, ``Camunda bpm platform loan assessment process lab,'' \emph{Brisbane, Australia: Queensland University of Technology}, 2013.

\bibitem{rademakers2012activiti}
T.~Rademakers, \emph{Activiti in Action: Executable business processes in BPMN 2.0}.\hskip 1em plus 0.5em minus 0.4em\relax Simon and Schuster, 2012.

\bibitem{MDAPW3}
\BIBentryALTinterwordspacing
A.~Samanipour, O.~Bushehrian, and G.~Robles, ``Mdapw3: Mda-based development of blockchain-enabled decentralized applications,'' \emph{Science of Computer Programming}, vol. 239, p. 103185, 2025. [Online]. Available: \url{https://www.sciencedirect.com/science/article/pii/S0167642324001084}
\BIBentrySTDinterwordspacing

\bibitem{omg_mda_guide}
\BIBentryALTinterwordspacing
{Object Management Group}, \emph{Model Driven Architecture (MDA) Guide}, \url{https://www.omg.org/mda/}, 2014, revision 2.0. [Online]. Available: \url{https://www.omg.org/mda/}
\BIBentrySTDinterwordspacing

\bibitem{corradini2021model_multiBlockchain}
F.~Corradini, A.~Marcelletti, A.~Morichetta, A.~Polini, B.~Re, E.~Scala, and F.~Tiezzi, ``Model-driven engineering for multi-party business processes on multiple blockchains,'' \emph{Blockchain: Research and Applications}, vol.~2, no.~3, p. 100018, 2021.

\bibitem{Traceability}
C.~Di~Ciccio, A.~Cecconi, J.~Mendling, and et~al., ``Blockchain-based traceability of inter-organisational business processes,'' in \emph{Business Modeling and Software Design}, B.~Shishkov, Ed.\hskip 1em plus 0.5em minus 0.4em\relax Cham: Springer International Publishing, 2018, pp. 56--68.

\bibitem{Dynamic-Integration}
P.~Klinger and F.~Bodendorf, \emph{Blockchain-based Cross-Organizational Execution Framework for Dynamic Integration of Process Collaborations}, 03 2020, pp. 893--908.

\bibitem{chor-chain}
\BIBentryALTinterwordspacing
F.~Corradini, A.~Marcelletti, A.~Morichetta, A.~Polini, B.~Re, and F.~Tiezzi, ``Engineering trustable and auditable choreography-based systems using blockchain,'' \emph{ACM Trans. Manage. Inf. Syst.}, vol.~13, no.~3, feb 2022. [Online]. Available: \url{https://doi.org/10.1145/3505225}
\BIBentrySTDinterwordspacing

\bibitem{Efficient_renda}
P.~Wang, Z.~Sun, R.~Li, J.~Chen, P.~Gong, and X.~Du, ``An efficient customized blockchain system for inter-organizational processes,'' in \emph{2023 IEEE International Conference on Web Services (ICWS)}, 2023, pp. 615--625.

\bibitem{Ethereum}
C.~Dannen, \emph{Introducing Ethereum and solidity}.\hskip 1em plus 0.5em minus 0.4em\relax Springer, 2017, vol.~1.

\bibitem{fabric}
\BIBentryALTinterwordspacing
E.~Androulaki, A.~Barger, V.~Bortnikov, and et~al., ``Hyperledger fabric: a distributed operating system for permissioned blockchains,'' in \emph{Proceedings of the Thirteenth EuroSys Conference}, ser. EuroSys '18.\hskip 1em plus 0.5em minus 0.4em\relax New York, NY, USA: Association for Computing Machinery, 2018. [Online]. Available: \url{https://doi.org/10.1145/3190508.3190538}
\BIBentrySTDinterwordspacing

\bibitem{bodorik2023_TABS}
P.~Bodorik, C.~G. Liu, and D.~Jutla, ``Tabs: Transforming automatically bpmn models into blockchain smart contracts,'' \emph{Blockchain: Research and Applications}, vol.~4, no.~1, p. 100115, 2023.

\bibitem{shenIcws}
X.~Shen, Z.~Wang, J.~Luo, H.~Ruan, H.~Xu, and M.~Liu, ``Ibc: An integrated framework combining blockchain with bpmn choreography to enhance multi-party collaboration,'' in \emph{2024 IEEE International Conference on Web Services (ICWS)}, 2024, pp. 457--467.

\bibitem{supplyExample}
W.~Fdhila, S.~Rinderle-Ma, D.~Knuplesch, and M.~Reichert, ``Change and compliance in collaborative processes,'' 06 2015, pp. 162--169.

\bibitem{firefly}
``{Hyperledger Firefly Documentation},'' \url{https://hyperledger.github.io/firefly/}, accessed: 2024-03-10.

\bibitem{cello}
``{Hyperledger Cello Home},'' \url{https://github.com/hyperledger/cello}, accessed: 2024-03-10.

\bibitem{Journal-software}
Z.~Fu-Li, H.~Pei-Yu, L.~Shan-Shan, L.~Shan-Shan, L.~Zhi-Ying, and D.~Meng-Jie, ``Framework for architecting smart contracts using microservices,'' \emph{Journal of Software}, vol.~32, no.~11, p. 3423, 11 2021.

\bibitem{benet2014ipfs}
J.~Benet, ``Ipfs-content addressed, versioned, p2p file system,'' \emph{arXiv preprint arXiv:1407.3561}, 2014.

\bibitem{oracle_pattern}
R.~Mühlberger, S.~Bachhofner, E.~Castelló~Ferrer, and et~al., ``Foundational oracle patterns: Connecting blockchain to the off-chain world,'' in \emph{Business Process Management: Blockchain and Robotic Process Automation Forum}, A.~Asatiani, J.~M. García, N.~Helander, and et~al., Eds.\hskip 1em plus 0.5em minus 0.4em\relax Cham: Springer International Publishing, 2020, pp. 35--51.

\bibitem{proctor2012drools}
M.~Proctor, ``Drools: a rule engine for complex event processing,'' in \emph{Applications of Graph Transformations with Industrial Relevance: 4th International Symposium, AGTIVE 2011, Budapest, Hungary, October 4-7, 2011, Revised Selected and Invited Papers 4}.\hskip 1em plus 0.5em minus 0.4em\relax Springer, 2012, pp. 2--2.

\bibitem{Caterpillar2017}
O.~Pintado, ``Caterpillar: {{A Blockchain-Based Business Process Management System}}.''

\bibitem{Caterpillar2019}
\BIBentryALTinterwordspacing
O.~López-Pintado, L.~García-Bañuelos, M.~Dumas, I.~Weber, and A.~Ponomarev, ``Caterpillar: {{A}} business process execution engine on the {{Ethereum}} blockchain,'' vol.~49, no.~7, pp. 1162--1193. [Online]. Available: \url{https://onlinelibrary.wiley.com/doi/abs/10.1002/spe.2702}
\BIBentrySTDinterwordspacing

\bibitem{Lorikeet}
\BIBentryALTinterwordspacing
A.~B. Tran, Q.~Lu, and I.~Weber, ``Lorikeet: A model-driven engineering tool for blockchain-based business process execution and asset management,'' in \emph{International Conference on Business Process Management}, 2018. [Online]. Available: \url{https://api.semanticscholar.org/CorpusID:52195200}
\BIBentrySTDinterwordspacing

\bibitem{LorikeetJournal}
\BIBentryALTinterwordspacing
Q.~Lu, A.~Binh~Tran, I.~Weber, and et~al., ``Integrated model-driven engineering of blockchain applications for business processes and asset management,'' \emph{Software: Practice and Experience}, vol.~51, no.~5, pp. 1059--1079, 2021. [Online]. Available: \url{https://onlinelibrary.wiley.com/doi/abs/10.1002/spe.2931}
\BIBentrySTDinterwordspacing

\bibitem{chor-chain2}
\BIBentryALTinterwordspacing
F.~Corradini, A.~Marcelletti, A.~Morichetta, A.~Polini, B.~Re, and F.~Tiezzi, ``Engineering trustable choreography-based systems using blockchain,'' in \emph{Proceedings of the 35th Annual ACM Symposium on Applied Computing}, ser. SAC '20.\hskip 1em plus 0.5em minus 0.4em\relax New York, NY, USA: Association for Computing Machinery, 2020, p. 1470–1479. [Online]. Available: \url{https://doi.org/10.1145/3341105.3373988}
\BIBentrySTDinterwordspacing

\bibitem{FlexChain1}
F.~Corradini, A.~Marcelletti, A.~Morichetta, and et~al., ``Flexible execution of multi-party business processes on blockchain,'' in \emph{2022 IEEE/ACM 5th International Workshop on Emerging Trends in Software Engineering for Blockchain (WETSEB)}, 2022, pp. 25--32.

\bibitem{FlexChain2}
\BIBentryALTinterwordspacing
F.~Corradini, A.~Marcelletti, A.~Morichetta, A.~Polini, B.~Re, and F.~Tiezzi, ``A flexible approach to multi-party business process execution on blockchain,'' \emph{Future Generation Computer Systems}, vol. 147, pp. 219--234, 2023. [Online]. Available: \url{https://www.sciencedirect.com/science/article/pii/S0167739X23001814}
\BIBentrySTDinterwordspacing

\bibitem{renmin-university}
P.~Wang, Z.~Sun, R.~Li, J.~Chen, P.~Gong, and X.~Du, ``An efficient customized blockchain system for inter-organizational processes,'' in \emph{2023 IEEE International Conference on Web Services (ICWS)}, 2023, pp. 615--625.

\bibitem{Rental-Claim}
J.~Ladleif, M.~Weske, and I.~Weber, ``Modeling and enforcing blockchain-based choreographies,'' in \emph{Business Process Management: 17th International Conference, BPM 2019, Vienna, Austria, September 1--6, 2019, Proceedings 17}.\hskip 1em plus 0.5em minus 0.4em\relax Springer, 2019, pp. 69--85.

\bibitem{Purchase}
I.~Compagnucci, F.~Corradini, F.~Fornari, and B.~Re, ``A study on the usage of the bpmn notation for designing process collaboration, choreography, and conversation models,'' \emph{Business \& Information Systems Engineering}, vol.~66, no.~1, pp. 43--66, 2024.

\bibitem{management-system}
P.~Sala, C.~Combi, M.~Mantovani, and R.~Rizzi, ``Discovering evolving temporal information: Theory and application to clinical databases,'' \emph{SN Computer Science}, vol.~1, no.~3, p. 153, 2020.

\bibitem{weber_survey}
\BIBentryALTinterwordspacing
J.~Mendling, I.~Weber, W.~V.~D. Aalst, and et~al., ``Blockchains for business process management - challenges and opportunities,'' \emph{ACM Trans. Manage. Inf. Syst.}, vol.~9, no.~1, feb 2018. [Online]. Available: \url{https://doi.org/10.1145/3183367}
\BIBentrySTDinterwordspacing

\bibitem{Blockchain-Support}
C.~Di~Ciccio, A.~Cecconi, M.~Dumas, and et~al., ``Blockchain support for collaborative business processes,'' \emph{Informatik Spektrum}, vol.~42, 06 2019.

\bibitem{Optimized}
L.~Garc{\'i}a-Ba{\~{n}}uelos, A.~Ponomarev, M.~Dumas, and I.~Weber, ``Optimized execution of business processes on blockchain,'' in \emph{Business Process Management}, J.~Carmona, G.~Engels, and A.~Kumar, Eds.\hskip 1em plus 0.5em minus 0.4em\relax Cham: Springer International Publishing, 2017, pp. 130--146.

\bibitem{Inter-organizational-by-blockchain}
H.~Nakamura, K.~Miyamoto, and M.~Kudo, ``Inter-organizational business processes managed by blockchain,'' in \emph{Web Information Systems Engineering -- WISE 2018}, H.~Hacid, W.~Cellary, H.~Wang, H.-Y. Paik, and R.~Zhou, Eds.\hskip 1em plus 0.5em minus 0.4em\relax Cham: Springer International Publishing, 2018, pp. 3--17.

\bibitem{BPMN-dmn-separation}
T.~Biard, A.~Le~Mauff, M.~Bigand, and J.-P. Bourey, ``Separation of decision modeling from business process modeling using new “decision model and notation”(dmn) for automating operational decision-making,'' in \emph{Risks and Resilience of Collaborative Networks}, L.~M. Camarinha-Matos, F.~B{\'e}naben, and W.~Picard, Eds.\hskip 1em plus 0.5em minus 0.4em\relax Cham: Springer International Publishing, 2015, pp. 489--496.

\bibitem{Real-world-DMN}
J.~Taylor and J.~Purchase, \emph{Real-World Decision Modeling with DMN}.\hskip 1em plus 0.5em minus 0.4em\relax USA: Meghan-Kiffer Press, 2016.

\bibitem{MDE4BBIS}
V.~A. de~Sousa and C.~Burnay, ``Mde4bbis: A framework to incorporate model-driven engineering in the development of blockchain-based information systems,'' in \emph{2021 Third International Conference on Blockchain Computing and Applications (BCCA)}, 2021, pp. 195--200.

\bibitem{DMN-execution-ethereum}
\BIBentryALTinterwordspacing
S.~Haarmann, K.~Batoulis, A.~Nikaj, and M.~Weske, ``Dmn decision execution on the ethereum blockchain,'' in \emph{Advanced Information Systems Engineering: 30th International Conference, CAiSE 2018, Tallinn, Estonia, June 11-15, 2018, Proceedings}.\hskip 1em plus 0.5em minus 0.4em\relax Berlin, Heidelberg: Springer-Verlag, 2018, p. 327–341. [Online]. Available: \url{https://doi.org/10.1007/978-3-319-91563-0\_20}
\BIBentrySTDinterwordspacing

\bibitem{execution-decision}
------, ``Executing collaborative decisions confidentially on blockchains,'' in \emph{Business Process Management: Blockchain and Central and Eastern Europe Forum}, C.~Di~Ciccio, R.~Gabryelczyk, L.~Garc{\'i}a-Ba{\~{n}}uelos, T.~Hernaus, R.~Hull, M.~Indihar~{\v{S}}temberger, A.~K{\H{o}}, and M.~Staples, Eds.\hskip 1em plus 0.5em minus 0.4em\relax Cham: Springer International Publishing, 2019, pp. 119--135.

\bibitem{Executing-DMN}
\BIBentryALTinterwordspacing
S.~Haarmann, \emph{Executing DMN Decisions on the Blockchain}.\hskip 1em plus 0.5em minus 0.4em\relax Cham: Springer International Publishing, 2021, pp. 43--53. [Online]. Available: \url{https://doi.org/10.1007/978-3-030-81409-0\_4}
\BIBentrySTDinterwordspacing

\end{thebibliography}
}
\vspace{-2cm}
\begin{IEEEbiography}[{\includegraphics[width=1in,height=1.25in,clip,keepaspectratio]{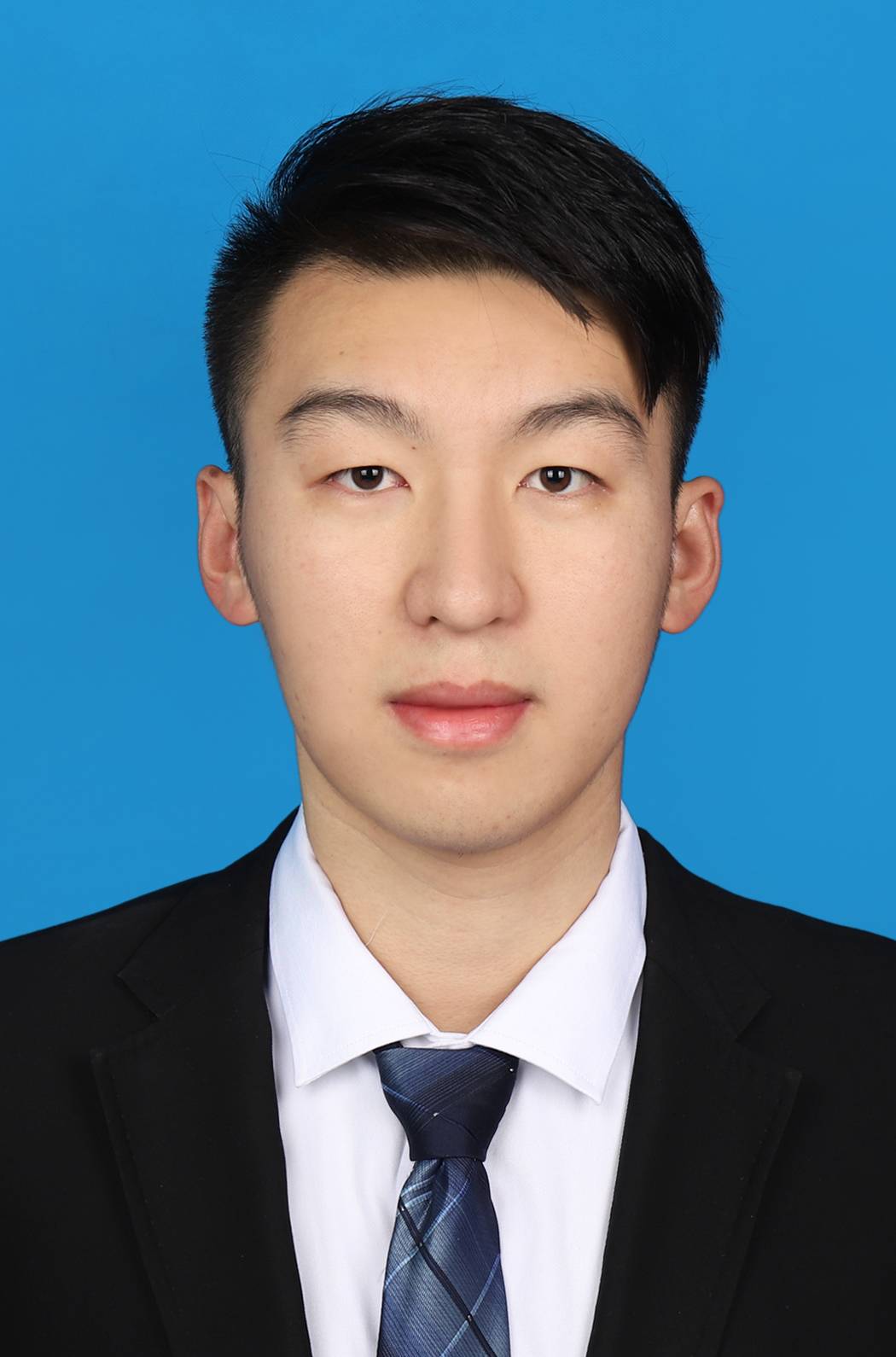}}]{Shen Xinzhe}Shen Xinzhe is a Ph.D. student in Software Engineering at Harbin Institute of Technology, under the supervision of Professor Wang Zhongjie. He is affiliated with the Enterprise Service and Intelligent Computing Research Center. His research focuses on trustworthy computing, blockchain, enterprise collaboration. 
\end{IEEEbiography}

\vspace{-1.4cm}
\begin{IEEEbiography}[{\includegraphics[width=1in,height=1.25in,clip,keepaspectratio]{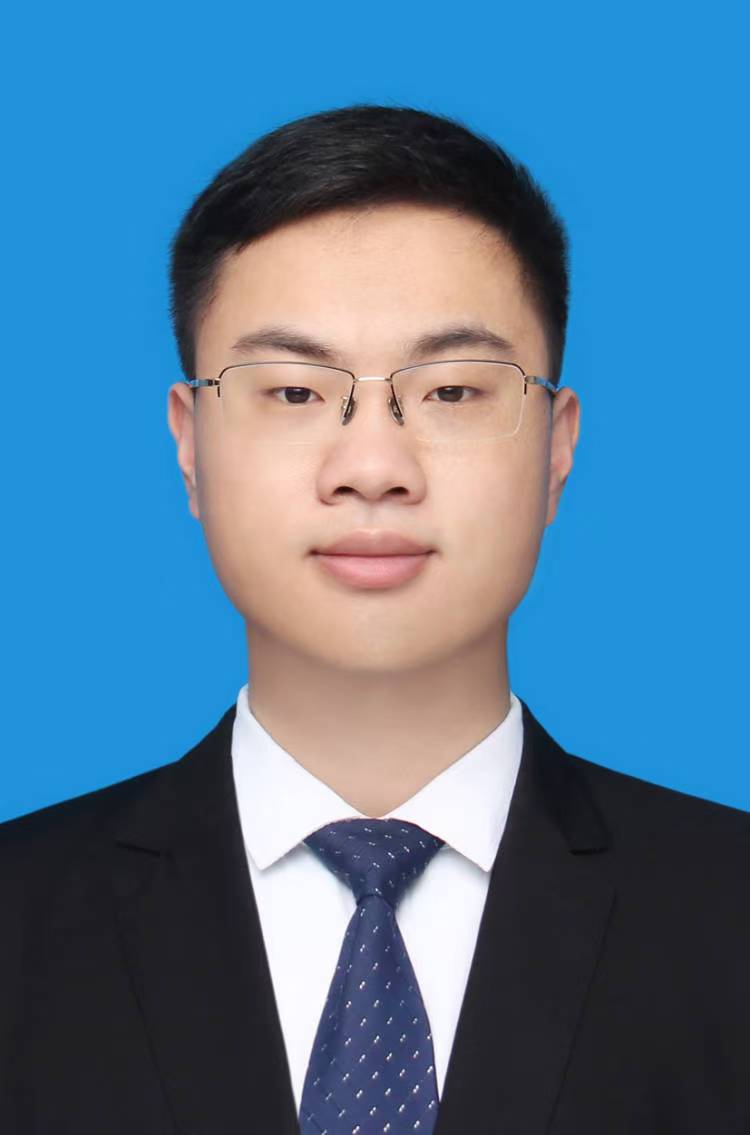}}]{Jiale Luo}Jiale Luo received the B.E. degree in Software Engineering from Harbin Engineering University in 2023. He is currently pursuing his M.S.degree in Software Engineering at Harbin Institute of Technology, China. His research focuses on trustworthy computing, blockchain, enterprise collaboration. 
\end{IEEEbiography}

\vspace{-1.4cm}
\begin{IEEEbiography}[{\includegraphics[width=1in,height=1.25in,clip,keepaspectratio]{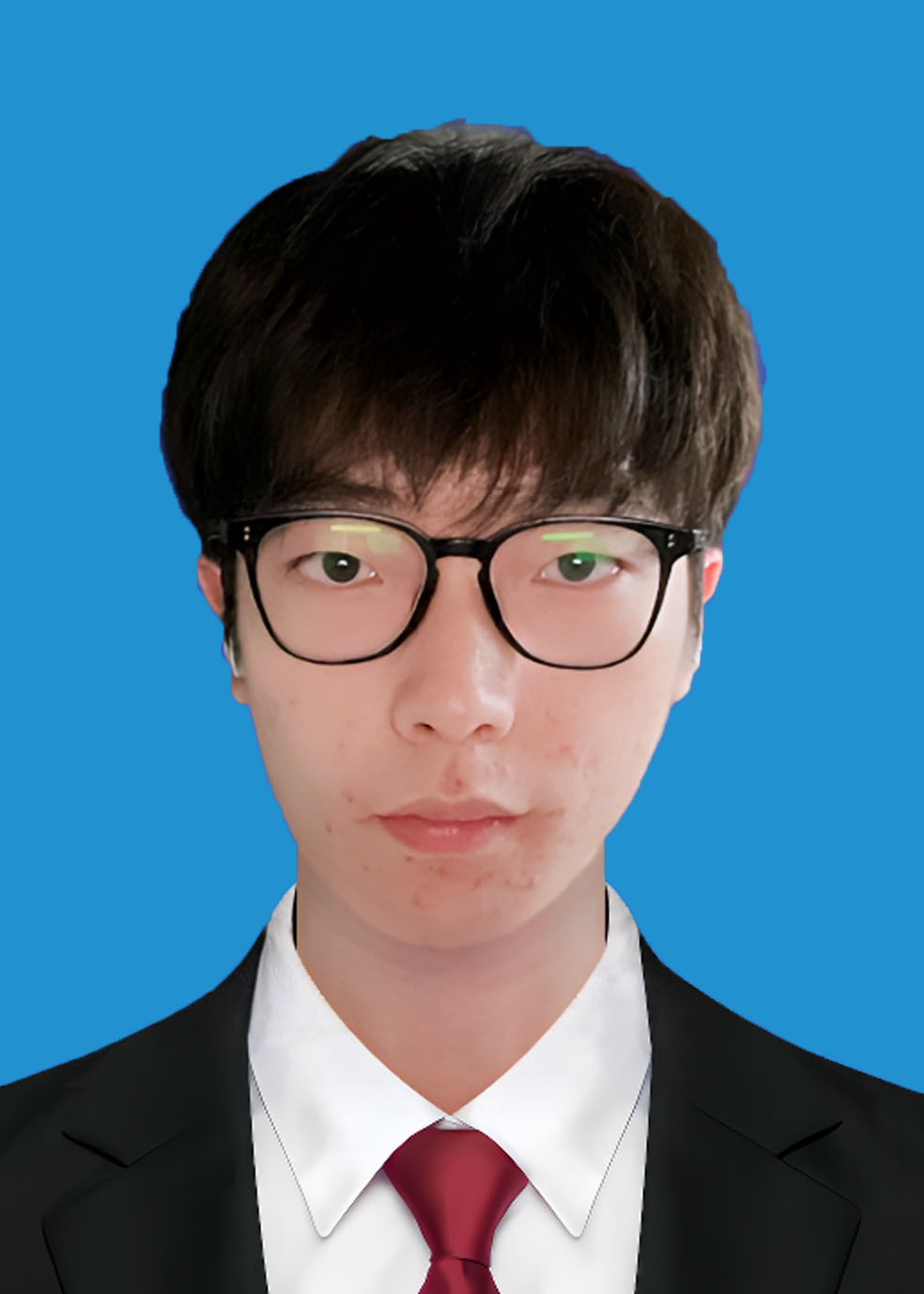}}]{Hao Wang}Hao Wang is currently pursuing his B.E. degree in Cyberspace Security at Harbin Institute of Technology, Weihai, China. His research focuses on blockchain, network security, cryptology. 
\end{IEEEbiography}

\vspace{-1.4cm}
\begin{IEEEbiography}[{\includegraphics[width=1in,height=1.25in,clip,keepaspectratio]{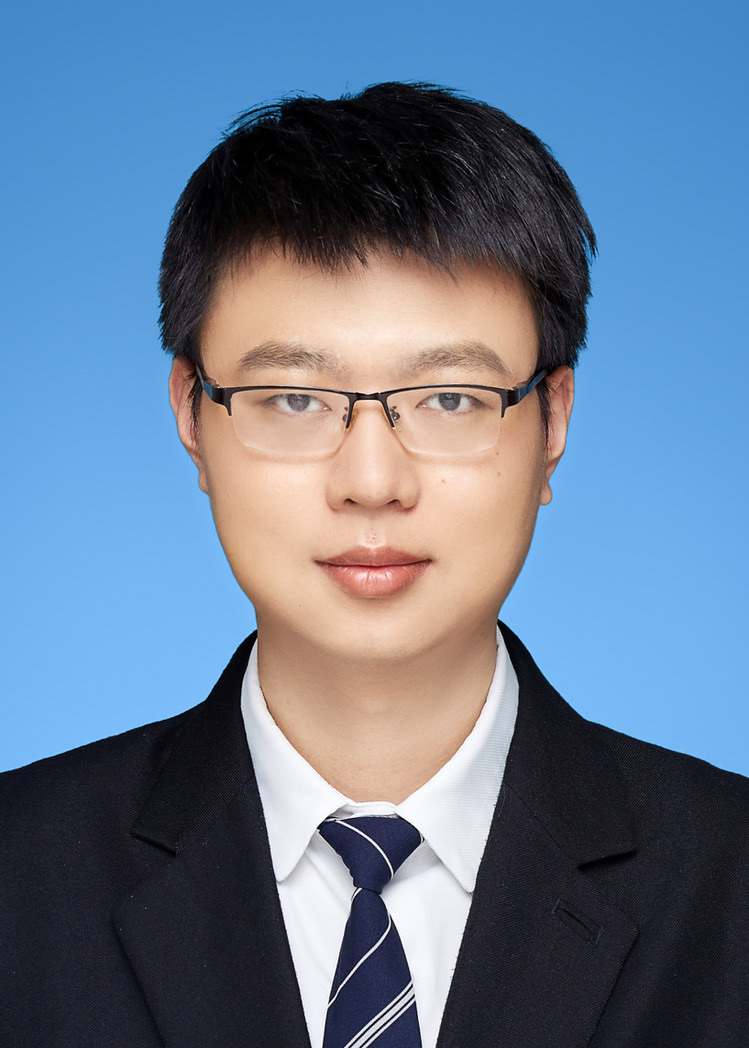}}]{Mingyi Liu} is an assistant professor at Faculty of Computing, Harbin Institute of Technology (HIT). He received the Ph.D. degree in software engineering from Harbin Institute of Technology in 2023. His research interests include service ecosystem model, service evolution analysis, data mining and graph neural networks. \end{IEEEbiography} 

\vspace{-1.4cm}
\begin{IEEEbiography}[{\includegraphics[width=1in,height=1.25in,clip,keepaspectratio]{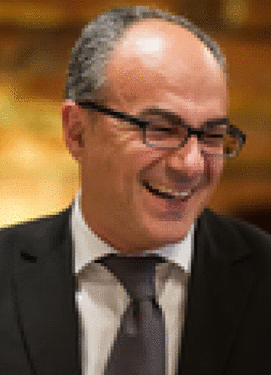}}]{Schahram Dustdar}(Fellow,IEEE) is a full professor of computer science (informatics) with a focus on Internet Technologies heading the Distributed Systems Group, TU Wien. He is chairman of the Informatics Section of the Academia Europaea (since December 9, 2016). He is a member of the IEEEConference Activities Committee (CAC) (since 2016), the Section Committee of Informatics of the Academia Europaea (since 2015), a member of the Academia Europaea: The Academy of Europe, Informatics Section (since 2013). He is the recipient of the ACM Distinguished Scientist Award (2009) and the IBM Faculty Award (2012). He is an associate editor of \textit{IEEE Transactions on Services Computing}, \textit{ACM Transactions on the Web}, and \textit{ACM Transactions on Internet Technology}, and on the editorial board of the \textit{IEEE Internet Computing}. He is the editor-in-chief of the \textit{Computing} (an SCI-ranked journal of Springer). \end{IEEEbiography}

\vspace{-1.4cm}
\begin{IEEEbiography}[{\includegraphics[width=1in,height=1.25in,clip,keepaspectratio]{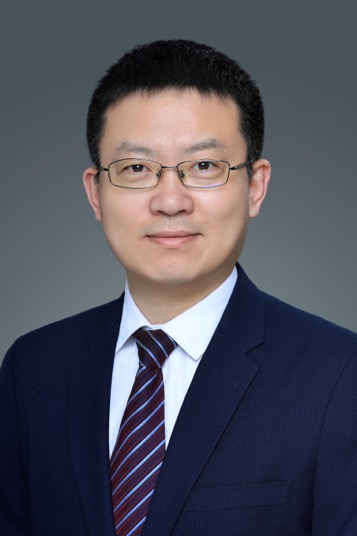}}]{Zhongjie Wang} is a professor and Director of Faculty of Computing, and Dean of School of Computer Science and Technology, Harbin Institute of Technology (HIT). He is a distinguished member of China Computer Federation (CCF), Associate Director of CCF Technical Committee of Services Computing and member of CCF Technical Committee of Software Engineering. His research interests include services computing, software engineering, cloud and edge computing, service governance, and service ecosystem evolution. He is a key member of a National Key Research and Development Plan project on industrial software. He has authored two books on services computing.
\end{IEEEbiography}
\end{document}